%% file: SUS-10-006_temp.tex
\pdfoutput=1

\documentclass[11pt,twoside,a4paper,cmspaper,final,collab]{cms-tdr}
\def\svnVersion{66706:66854}\def\cmsCernNo{2011-084}\def\cmsCernDate{\today}\def\cmsMessage{Submitted to the Journal of High Energy Physics}

\begin{document}\cmsNoteHeader{SUS-10-006}

\hyphenation{had-ron-i-za-tion}
\hyphenation{cal-or-i-me-ter}
\hyphenation{de-vices}
\RCS$Revision: 66854 $
\RCS$HeadURL: svn+ssh://alverson@svn.cern.ch/reps/tdr2/papers/SUS-10-006/trunk/SUS-10-006.tex $
\RCS$Id: SUS-10-006.tex 66854 2011-07-10 16:14:55Z alverson $
\def\eslash{{\hbox{$E$\kern-0.6em\lower-.05ex\hbox{/}\kern0.10em}}}
\def\met{\mbox{$\eslash_\text{T}$}\xspace}
\def\beslash{{\hbox{$E$\kern-0.4em\lower-.05ex\hbox{/}\kern0.10em}}}
\def\bmet{\mbox{$\beslash_\text{T}$}\xspace}
\newcommand{\sMET}{\ensuremath{Y_{\rm MET}}\xspace}
\newcommand{\cPK}{\ensuremath{\cmsSymbolFace{K}}\xspace}
\cmsNoteHeader{SUS-10-006} 
\title{Search for supersymmetry in pp collisions at $\sqrt s =7$ TeV in events with a single lepton, jets, and missing transverse
momentum}%

\date{\today}

\abstract{
Results are reported from a search for physics beyond the standard
model in proton-proton collisions at a center-of-mass energy of 7 TeV,
focusing on the signature with a single, isolated, high-transverse-momentum lepton (electron
or muon), energetic jets, and
large missing transverse momentum. The data sample comprises an
integrated luminosity of 36 pb$^{-1},$ recorded by
the CMS experiment at the LHC. The search
is motivated by models of new physics, including supersymmetry.
The observed event yields
are consistent with standard model backgrounds predicted using
control samples obtained from the data. The characteristics of the
event sample are consistent with those expected for
the production of \cPqt\cPaqt\ and \PW +jets events.
The results are interpreted in terms of limits on the
parameter space for the constrained minimal supersymmetric
extension of the standard model.}

\hypersetup{%
pdfauthor={CMS Collaboration},%
pdftitle={Search for supersymmetry in pp collisions at sqrt(s)=7 TeV in events with a single lepton, jets, and missing transverse momentum},%
pdfsubject={CMS},%
pdfkeywords={CMS, physics, supersymmetry, SUSY}}

\maketitle 

\input{Introduction_Final}

\input{EventSelection_Final}

\input{BackgroundEstimation_Final}
\input{Results_Final}

\input{Conclusions_Final}

\input{Acknowledgments}
\bibliography{auto_generated}   
\cleardoublepage \appendix\section{The CMS Collaboration \label{app:collab}}\begin{sloppypar}\hyphenpenalty=5000\widowpenalty=500\clubpenalty=5000\input{SUS-10-006-authorlist.tex}\end{sloppypar}
\end{document}

%% file: Introduction_Final.tex
\section{Introduction}
\label{sec:Introduction}

Searches for new physics at the TeV energy scale are motivated by 
several considerations, ranging from the strong astrophysical evidence
for dark matter~\cite{ref:Zwicky,ref:Trimble,ref:Bartelmann,ref:Feng} to theoretical issues associated with explaining
the observed particle masses and their hierarchy~\cite{ref:hierarchy1,ref:hierarchy2}. 
In this paper, we report results from a search for new physics in 
proton-proton collisions at a center-of-mass energy of 7~TeV, focusing on the 
signature with a single isolated lepton (electron or muon), 
multiple energetic jets, and large missing momentum transverse 
to the beam direction ($\met$). The data sample was collected by the Compact Muon Solenoid (CMS) 
experiment during 2010 at the Large Hadron Collider (LHC)
and corresponds to an integrated luminosity of $36$ pb$^{-1}$~\cite{ref:lumipas}.

The search signature arises naturally 
in several theoretical frameworks for new physics, among them 
supersymmetry (SUSY)~\cite{ref:Martin,ref:SUSY0,ref:SUSY1,ref:SUSY2,ref:SUSY3,ref:SUSY4}.
SUSY models predict a spectrum of new particles with couplings identical
to those of the standard model (SM), but with spins differing by half a unit
with respect to their SM partners. 
In many models, a multiplicatively conserved quantum number, $R$ parity, 
is introduced, constraining SUSY particles to be produced in pairs and
SUSY particle decay chains to end 
with the lightest supersymmetric particle (LSP). 
In some scenarios, the LSP is a neutralino, a heavy, 
electrically neutral, weakly interacting particle with the characteristics required 
of a dark-matter candidate. 

Searches at the Tevatron~\cite{CDFLimits,D0Limits,Abazov200934} 
and LEP~\cite{LEPLimits,aleph,delphi,l3,opal} have found no evidence as yet
for SUSY particles, demonstrating that, if supersymmetry exists, it is
broken, with SUSY particle masses typically greater than 100--300~GeV. 
Recently, searches from the 
CMS~\cite{ref:CMSalphaT,ref:SUS-10-002,ref:SUS-10-004,ref:SUS-10-007,ref:SUS-11-002,ref:SUS-10-008} 
and ATLAS~\cite{ref:ATLASSingleLepton,ref:ATLASSquarks,ref:ATLASBjets,ref:ATLASLeptonPairs,ref:ATLASOSLeptonPairs} 
experiments have extended the sensitivity to higher mass scales. 
In particular, ATLAS has reported~\cite{ref:ATLASSingleLepton} 
constraints on SUSY models from a search in the single-lepton channel,
which is defined in a similar manner to this analysis. 

At the LHC, relatively large cross sections for SUSY particle production (up
to tens of pb) can arise from strong-interaction (QCD) processes leading
to the production of gluino-gluino, squark-gluino, squark-squark, 
and squark-antisquark pairs. The search signature reflects the complex 
decay chains of the heavy, strongly coupled SUSY particles. 
The isolated lepton indicates a weak decay of a heavy particle, 
either a \PW\  boson or a new particle. Large missing momentum transverse to 
the beam direction can be carried by a neutrino or, in the case
of new physics, by one or more heavy, weakly 
interacting particles, such as the LSP. Finally, multiple jets can arise
from quarks and gluons produced in the decay chains. 
This signature arises in many SUSY models, including the constrained
minimal supersymmetric extension to the standard model (CMSSM)~\cite{ref:CMSSM,ref:MSUGRA}, which
we use to interpret the results.

The SUSY signal is not characterized by any narrow peaks, but rather
by broad distributions that extend to higher values of the kinematic variables 
than those of the SM backgrounds. These backgrounds 
arise primarily from the production of \cPqt\cPaqt, \PW +jets, and QCD multijet events. 
It is therefore critical to determine the 
extent of the tails of the SM background
distributions. We use methods that are primarily based
on control samples in the data, sometimes in conjunction 
with certain reliable information from simulated event samples. 

Two complementary methods are used to probe the event sample, one focusing 
mainly on jets and $\met$, and the other emphasizing the lepton transverse momentum (\pt)  
and $\met$.  The first method uses two
kinematic variables, \HT and $\met/\sqrt{\HT}$, where \HT is 
the scalar sum of the jet \pt  values for all jets
above a certain threshold. The yields in three control 
regions are combined to provide a prediction for the total background in the signal region, without
differentiating among the backgrounds. This method tests whether the
behavior of the event sample with respect to jets and $\met$
is consistent with that expected from the SM.

The second method, which is ultimately used for the interpretation in terms of constraints
on SUSY parameter space, exploits the relationship between the lepton \pt  and the $\met$ 
distributions. The dominant SM backgrounds are \cPqt\cPaqt\ and \PW $(\ell\nu)$+jets events 
with a single, isolated, high-\pt  lepton ($\ell=\Pe$ or $\mu$). In these processes, the lepton \pt  and
the $\met$ distributions are closely related, because the lepton and neutrino 
are produced together in the two-body \PW\  decay.
The observed lepton spectrum, with appropriate corrections, can therefore 
be used to predict the $\met$ 
spectrum under the null (SM) hypothesis. In contrast, the distributions of 
lepton \pt  and $\met$ are very different in many SUSY models, where the presence of two
LSPs effectively decouples the two distributions.
In such models, the method is robust against potential signal contamination of the control regions.
Smaller backgrounds from dilepton \cPqt\cPaqt\ events (where both \PW\ 
bosons associated with the top quarks decay leptonically) feeding down to the
single-lepton channel, and from $\tau\to\ell$ decays in both \cPqt\cPaqt\ and \PW +jets events, are 
estimated from additional control samples, 
as is the QCD multijet background.

The two methods for probing the data provide a broader picture of the event sample than
a single approach. Given the large range of potential signal models, the use of 
multiple methods for the background determination provides valuable information 
to ensure that the event sample is comprehensively understood.

The CMS detector, described in detail in Ref.~\cite{ref:CMS}, is a multipurpose
apparatus designed to study high-\pt  physics processes in proton-proton collisions,
as well as a broad range of phenomena in heavy-ion collisions. 
The central element of CMS is a 3.8 T superconducting solenoid, 
13 m in length and 6 m in diameter. Within the magnet are (in order
of increasing radius from the beam pipe) the high-precision silicon-pixel
and silicon-strip detectors for charged particle tracking;
a lead tungstate crystal electromagnetic calorimeter for 
measurements of photons, electrons, and the electromagnetic component
of jets; and a hadron calorimeter, constructed from scintillating tiles and brass
absorbers, for jet-energy measurements.
Beyond the magnet is the muon system, comprising drift-tube, cathode-strip, 
and resistive-plate detectors interleaved with steel absorbers. Each detector
system comprises subsystems that cover the central (barrel)
and forward (endcap) regions.

In describing the angular distribution of particles and the acceptance of the detector, we 
frequently make use of the pseudorapidity, $\eta = -\ln[\tan(\theta/2)]$, where 
the polar angle $\theta$ of the particle's momentum vector is measured
with respect to the $z$ axis of the CMS coordinate system. The $z$ axis
points along the direction of the counterclockwise rotating beam; the azimuthal
angle $\phi$ is measured in a plane perpendicular to this axis. The separation
between two momentum vectors in $\eta$-$\phi$ space is characterized by the
quantity $\Delta R = \sqrt{(\Delta\eta)^2+ (\Delta\phi)^2}$, which is approximately
invariant under Lorentz boosts along the $z$ axis.  

The paper is organized as follows. 
The event selection requirements are described in Section~\ref{sec:EventSelection}. 
Section~\ref{sec:BackgroundDetermination} begins with
a brief survey of the kinematic distributions, comparing the
data with simulated Monte Carlo (MC) event samples. The methodologies
for obtaining SM background estimates from control samples in the
data are described, and the observed yields in the data are compared
with these estimates. The systematic uncertainties are summarized in Section~\ref{sec:SysErr}.
Finally, the results, interpretation, and conclusions of the analysis are 
presented in Sections~\ref{sec:Results} 
and \ref{sec:Conclusions}.

%% file: EventSelection_Final.tex
\section{Event Samples and Preselection} 
\label{sec:EventSelection}
This section describes the overall strategy of the analysis, the event samples used, 
and the preselection requirements. The composition of the event sample is determined
largely by the topological requirements of a single isolated, high-\pt  lepton, 
either an electron or a muon, and at least four jets. The lepton-isolation requirement is 
critical for the rejection of QCD multijet processes, which have very large
cross sections. While many lepton candidates are produced in the semileptonic 
decays of \cPqb\  and \cPqc\  hadrons, from $\pi$ and \cPK\ decays in flight, and 
from misidentification of hadrons, the vast majority
of these are embedded in hadronic jets and are rejected using the lepton-isolation
variable described below. The initially very large \PW +jets background (which is 
dominated by \PW~$\to \Pe\nu$ or \PW~$\to\mu\nu$) 
is heavily suppressed by the four-jet requirement; \cPqt\cPaqt\ then emerges 
as the largest contribution to the background in the sample of events with moderate 
to large values of missing transverse momentum (above approximately 150~GeV). 

Because the analysis is part of a broad set of CMS topological SUSY searches involving $\met$, we veto events
containing a second isolated-lepton candidate. This procedure
reduces the statistical overlap between the searches in different topologies, provides a clearer
phenomenological interpretation of each search, and, in the single-lepton channel, suppresses SM backgrounds
that produce two or more isolated leptons. Nevertheless, \cPqt\cPaqt\ backgrounds with dileptons
can still feed into the sample, and this contribution must be determined, particularly because
the presence of two neutrinos can result in large values of \met. The background involving
\PW~$\to \tau\nu$ decays, both from \cPqt\cPaqt\ events and from direct \PW\  production, must
also be determined.

The analysis procedures are designed by studying simulated event samples based on
a variety of generators; in all cases except for certain SUSY scans discussed later, 
the detector simulation is performed using the 
\GEANTfour\ package~\cite{GEANT4}. QCD samples are generated with the 
\PYTHIA6.4.22~\cite{pythia} MC generator with tune Z2~\cite{ref:TuneZ2}. The dominant background,
\cPqt\cPaqt, is studied with a sample generated with \MADGRAPH 4.4.12~\cite{madgraph}. 
The \PW +jets and \cPZ +jets processes
are simulated with both \MADGRAPH and \ALPGEN~\cite{ALPGEN}. 

SUSY benchmark models are generated with \PYTHIA. Two models, designated LM0
and LM1 \cite{PTDR2}, are frequently used in CMS because they have large cross sections and are 
accessible with small event samples. LM0 is described by the   
universal scalar mass parameter
$m_0=200$~GeV, the universal gaugino mass parameter $m_{1/2}=160$~GeV, the universal
trilinear soft SUSY breaking parameter $A_0=-400$~GeV, 
the ratio of the two Higgs-doublet vacuum expectation values $\tan\beta=10$, and the sign of
the Higgs mixing parameter
$\mu>0$. For LM1, the corresponding parameters are $m_0=60$~GeV, $m_{1/2}=250$~GeV, $A_0=0$~GeV, 
$\tan\beta=10$, and $\mu>0$. The leading order cross sections for these
models are 38.9 pb (LM0) and 4.9 pb (LM1); with K factors averaged over 
the contributing subprocesses, the next-to-leading order cross sections
are approximately 54.9 pb (LM0) and 6.6 pb (LM1). These benchmark
models are beyond the exclusion limits of the Tevatron and LEP searches
referenced in Section~\ref{sec:Introduction}, and have recently been excluded by
LHC searches, e.g., Ref.~\cite{ref:CMSalphaT}. They provide useful comparison
points for searches in different channels. 
We also perform scans over CMSSM parameter space
using a large number of Monte Carlo samples in which the simulation is performed using
a CMS fast simulation package to reduce the time associated with the detector simulation.

The data samples used in the analysis are recorded using trigger
paths that directly require 
the presence of a lepton above a minimum \pt  threshold, sometimes in 
conjunction with additional jet energy. The basic muon trigger path
is a simple, single-muon trigger requiring $\pt(\mu)>11$~GeV. 
As the LHC luminosity increased above $2\times 10^{32}$~cm$^{-2}$s$^{-1}$, 
a trigger was implemented requiring both $\pt(\mu)> 5$ GeV and
$\HT^{\rm trigger}>70$~GeV, where $\HT^{\rm trigger}$ is 
the scalar sum of the raw calorimeter jet \et values measured at the trigger level. 
For electrons, a higher single-electron trigger threshold is 
required, $\pt(\Pe)>17$~GeV.

The offline preselection requirements are designed to be simple
and robust. Events are required to have at least one good 
reconstructed primary vertex, at least four jets, 
and exactly one isolated muon or exactly one isolated electron. (The
jet and lepton selection criteria are specified below.) 
The primary vertex must satisfy a set of quality requirements,
including $|z_{\rm PV}|<24$ cm and $\rho_{\rm PV}<2$ cm, where $z_{\rm PV}$ and $\rho_{\rm PV}$ are
the longitudinal and transverse distances of the primary vertex 
with respect to the nominal CMS interaction point.  

Jets and $\met$ are reconstructed using a particle-flow algorithm~\cite{ref:PAS-PFT-09-001,ref:PAS-PFT-10-002},
which combines information from all components of the detector. The $\met$ vector 
is defined as the negative of the vector sum of the transverse momenta of all
the particles reconstructed and identified by the particle-flow algorithm.
(The $\met$ quantity itself is the magnitude of the $\met$ vector.) 
The jet clustering is performed using the 
anti-$k_{\mathrm T}$ clustering algorithm~\cite{ref:antikt} with a distance parameter
of 0.5. Corrections based on simulation are applied 
to the raw jet energies to establish
a relative uniform response across the detector in $\eta$ and 
an absolute calibrated response in \pt . Additional jet-energy
corrections are applied to the data to take into account residual 
differences between the jet-energy calibration in data and
simulation. 
The performance of CMS jet reconstruction and the corrections are described in
Refs.~\cite{ref:PAS-JME-10-003,ref:PAS-JME-10-010}. Jet candidates are required to 
satisfy quality criteria that suppress noise and spurious
energy deposits, and each event must contain at 
least four jets with $\pt>30$ GeV and $|\eta|< 2.4$. 

In the muon channel, the preselection requires 
a single muon candidate~\cite{ref:PAS-MUO-10-002} satisfying
$\pt(\mu)>15$~GeV and $|\eta|<2.1$.
Several requirements are imposed on the elements that form the muon candidate.
The reconstructed track must have at least 11 hits in the silicon tracker, with an 
impact parameter $d_0$ in the transverse plane with respect to the beam spot 
satisfying $d_0<0.02$~cm and an impact parameter $d_z$ with respect to the 
primary vertex along the $z$ (beam) direction satisfying
$|d_z|<1.0$~cm. To suppress background in which the muon
originates from a semileptonic decay of a \cPqb\  or \cPqc\  quark in a jet, we 
require that it be spatially isolated from other energy in the event. A
cone of size $\Delta R=0.3$ is constructed around the muon
direction in $\eta$-$\phi$ space. The muon isolation variable, 
$I=\sum_{\Delta R<0.3}(\et + \pt)$, is defined as the sum
of the transverse energy \et (as measured in the electromagnetic and hadron
calorimeters) and the transverse
momentum \pt  (as measured in the silicon tracker) of all reconstructed objects 
within this cone, excluding the muon. This quantity is used to compute
the isolation relative to the muon transverse momentum, which is required to satisfy
$I/\pt(\mu)<0.1$. Finally,
the muon must satisfy $\Delta R>0.3$ with 
respect to all jets with $\pt>30$ GeV and $|\eta|<2.4$. 

For the electron channel, a single electron candidate~\cite{ref:PAS-EGM-10-004} is required to 
satisfy $\pt>20$ GeV and $|\eta|<2.4$, excluding the barrel-endcap overlap
region $1.44<|\eta|<1.57$. The relative isolation variable, defined as in the muon case, must
satisfy $I/\pt(\Pe)<0.07$ in  the barrel region and $I/\pt(\Pe)<0.06$ in the
endcaps, as well as a set of quality and photon-conversion rejection criteria. 
Events with two or more good lepton candidates are rejected, for the reasons discussed above.

%% file: BackgroundEstimation_Final.tex
\section{Signal Regions and Background Determination}
 \label{sec:BackgroundDetermination}

We next survey the properties of the event sample after imposing the 
preselection requirements described in the previous section, and 
after further requiring $\met>25$ GeV. While this $\met$ requirement is far
looser than that used in the final selection, it nevertheless
suppresses much of the remaining QCD multijet background and brings the sample 
closer to the final composition dominated by \cPqt\cPaqt\ and \PW +jets events.
The overall shapes of the observed distributions are found to be consistent
with those expected for these backgrounds. We then proceed
to apply methods, based on control samples in the data, 
that are designed to determine the SM contributions to the tails of the 
kinematic distributions in the signal regions. 

The quantity \HT
is defined as the scalar sum of the transverse momenta of jets $j$ with
$\pt^j>20$ GeV and $|\eta|<2.4$,
\begin{eqnarray}
\HT = \sum_{j}\pt^{j}.
\label{eq:HTdef}
\end{eqnarray}
A simple requirement on the \pt of the highest \pt or the two highest \pt jets 
can also provide discrimination between signal and background,
but such a requirement is more directly sensitive to the mass splittings in a 
new physics model than \HT. Thus, we prefer to use \HT to reduce the 
potential model dependence of the analysis. 

\begin{figure}[tb!]
 \begin{center}
\includegraphics[angle=0,width=0.32\textwidth]{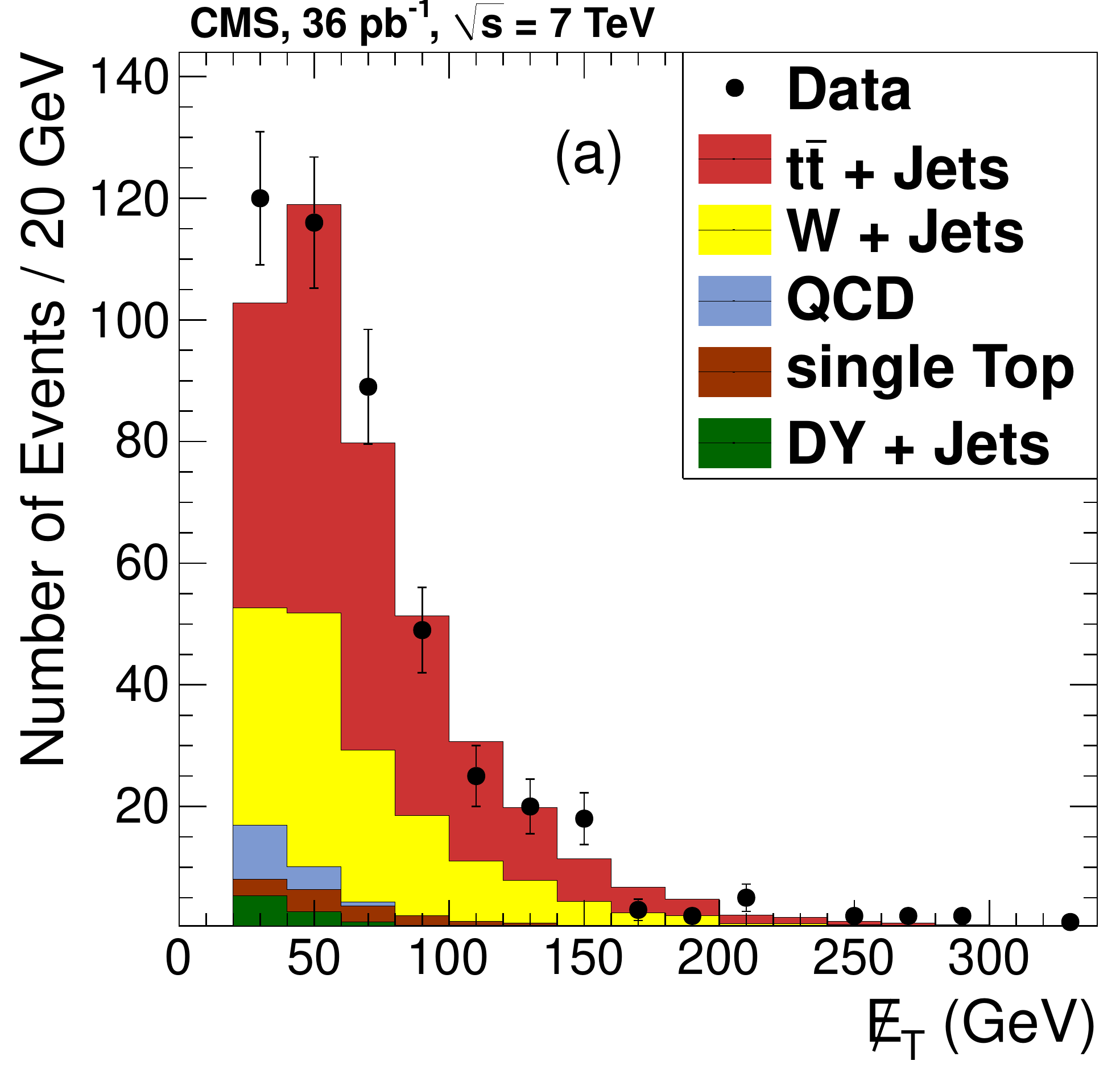}
\includegraphics[angle=0,width=0.32\textwidth]{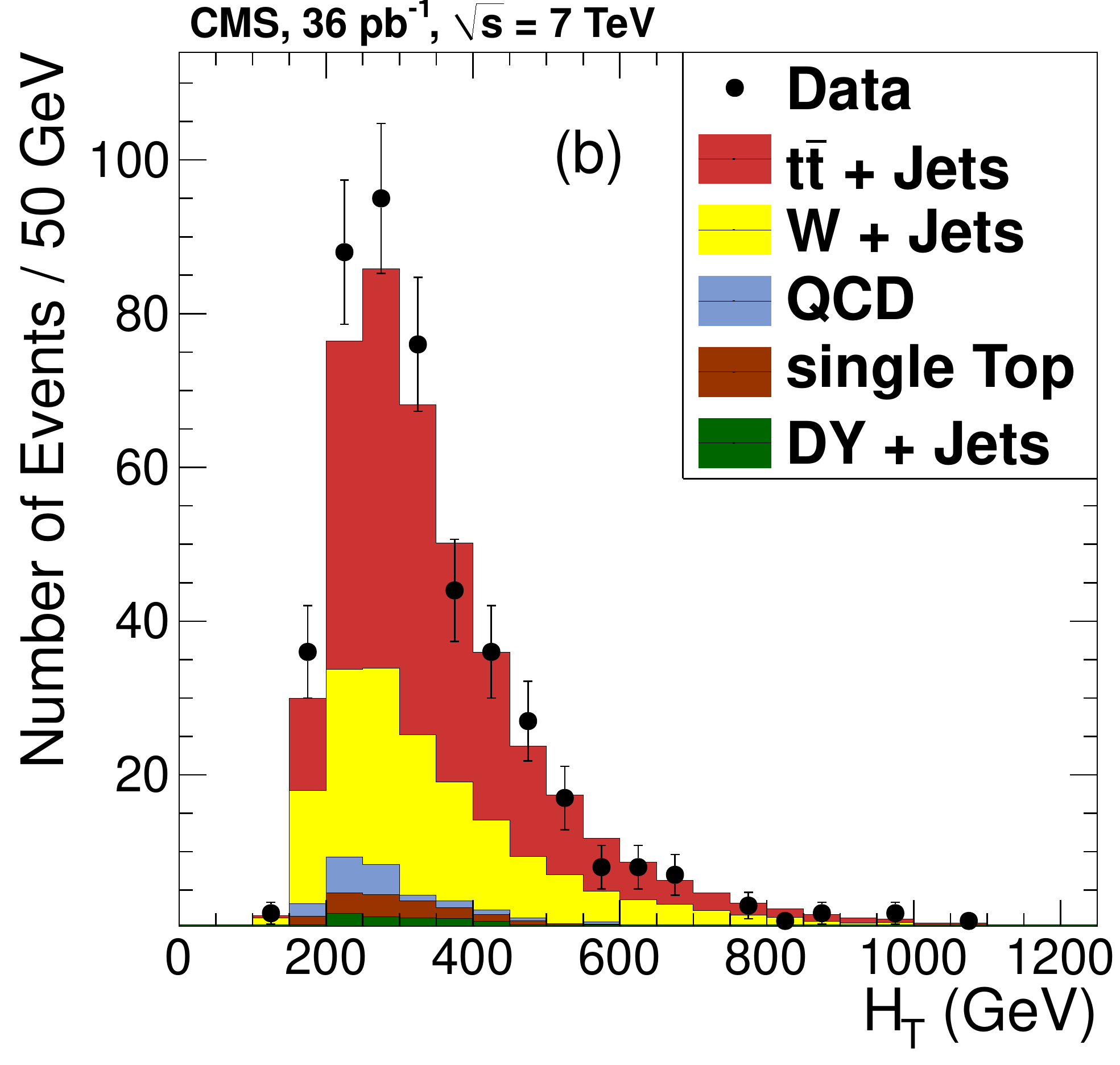}
\includegraphics[angle=0,width=0.32\textwidth]{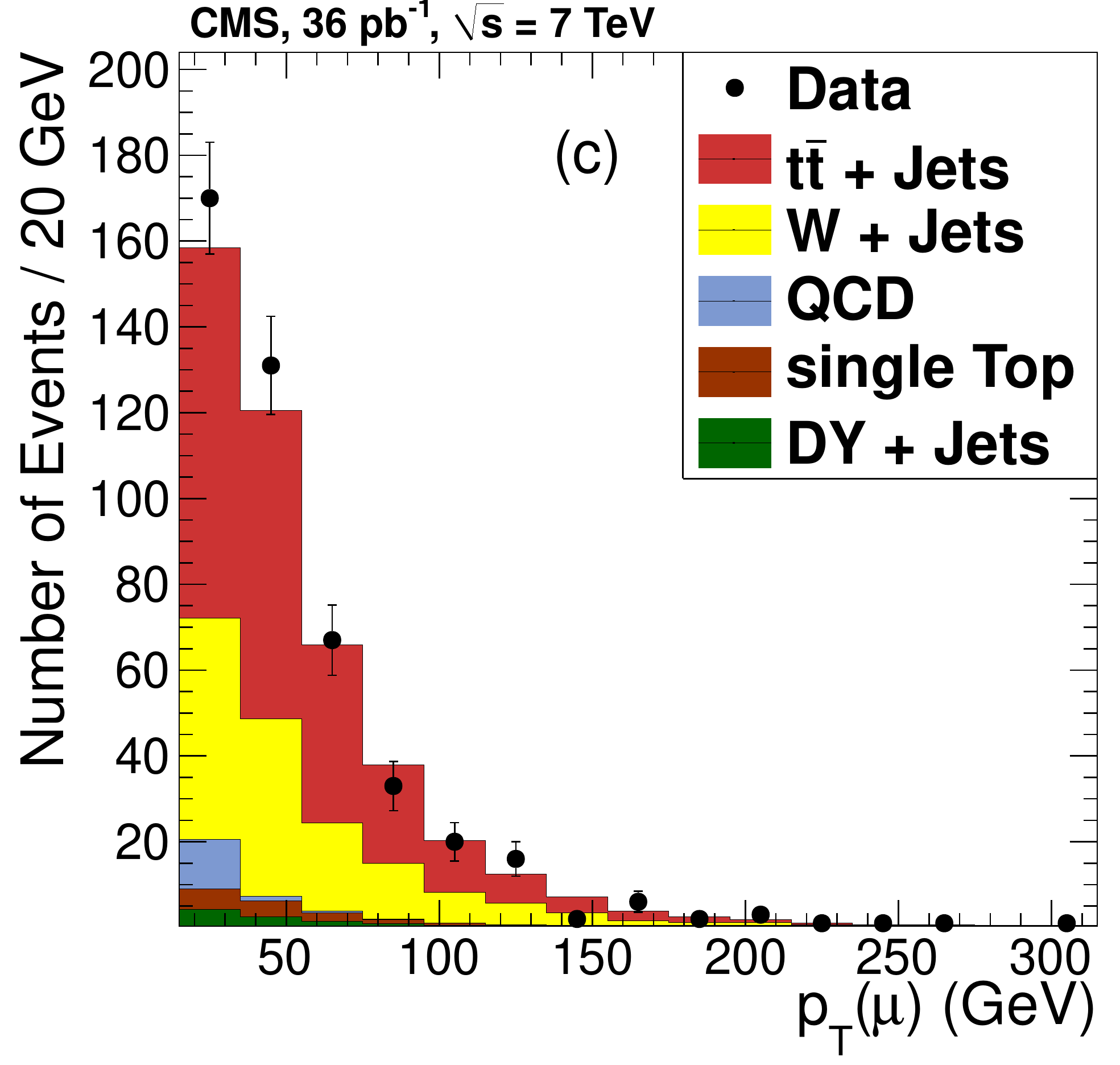} \\
\includegraphics[angle=0,width=0.32\textwidth]{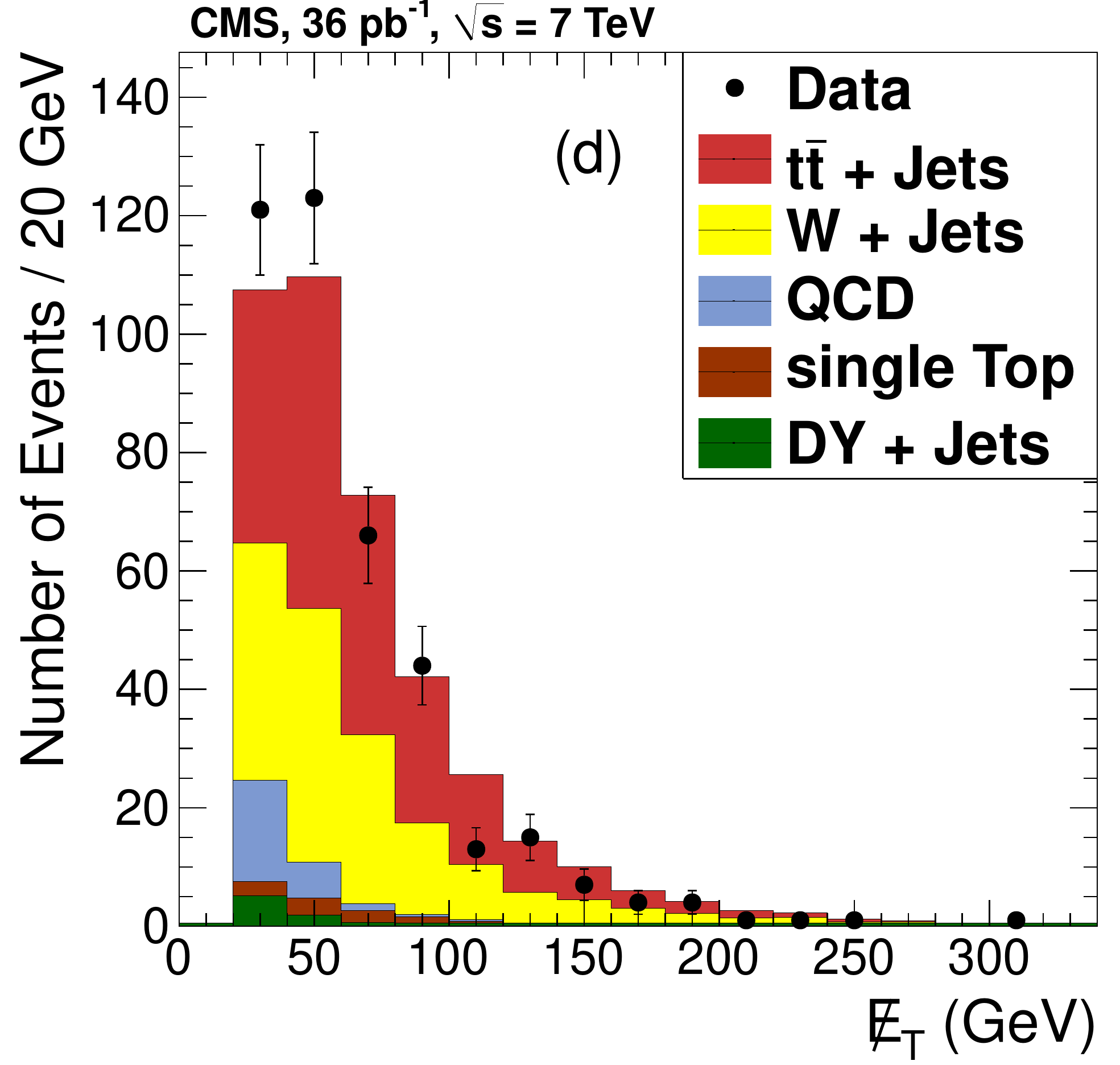}
\includegraphics[angle=0,width=0.32\textwidth]{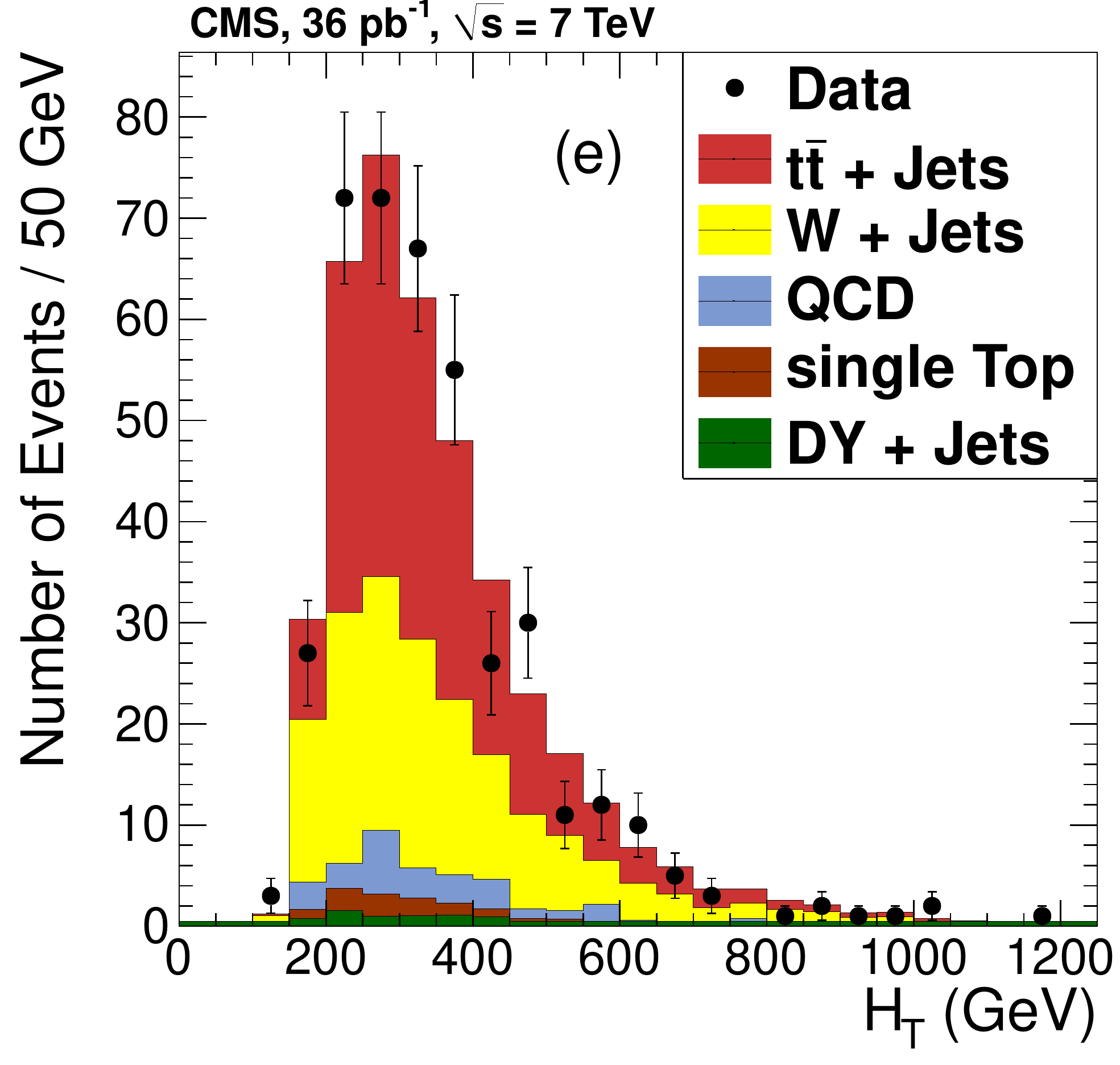}
\includegraphics[angle=0,width=0.32\textwidth]{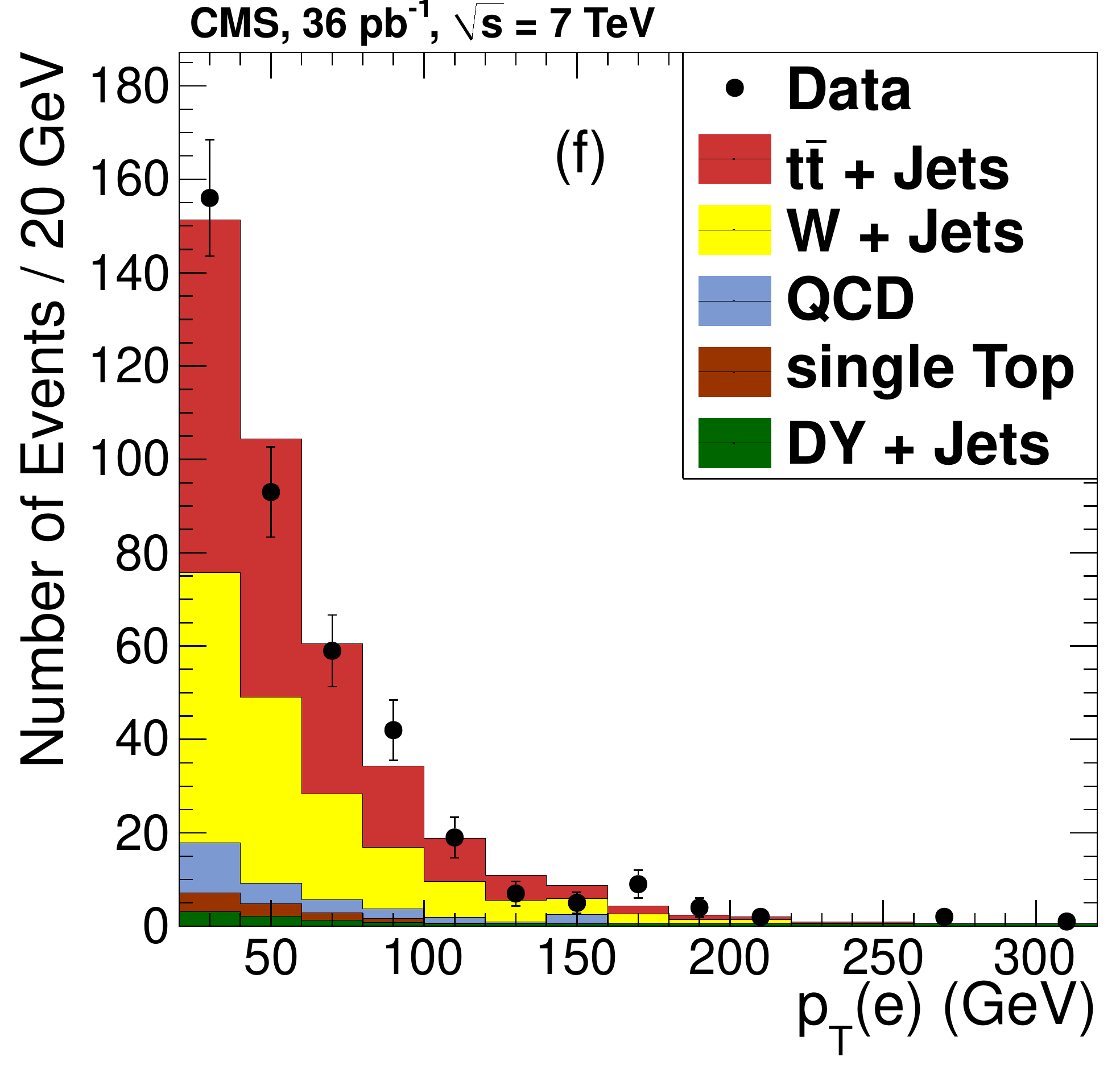}
 \end{center}
\caption{Comparison of distributions in data and simulated event samples for the muon channel: (a) $\met$, (b) \HT, and (c) lepton \pt; 
and for the electron channel (d) $\met$, (e) \HT, and (f) lepton \pt .
The data are shown by points with error bars; the 
simulation is displayed as the histogram with the individual components stacked. 
The preselection and $\met>25$ GeV requirements have been applied. 
}
\label{fig:EventProperties}
\end{figure}

After applying the preselection and $\met>25$ GeV as a loose requirement, 
444 (391) events are observed in data in the muon (electron) channel, compared with
395 (327) muon (electron) events in the simulated event samples. 
The estimate from simulation is based on summing the yields
from \cPqt\cPaqt, \PW+jets, Drell-Yan/\cPZ+jets, QCD multijet, and single-top production.
The contributions from \cPqt\cPaqt\ and \PW+jets account for about 90\% 
of the predicted yield. The number of events obtained from simulation is not
the basis for our background predictions, which rely primarily on control 
samples in the data. However, the approximate agreement between yields in 
data and simulation is a first indication that the analysis methods 
will be applied to a sample dominated by SM events.

Figure~\ref{fig:EventProperties} shows the distributions from data and simulated event samples
of three fundamental quantities in the muon and electron 
channels: \met, \HT, and lepton \pt . 
For the purpose of these comparisons, the normalization of the \cPqt\cPaqt\ sample 
is fixed by
the integrated luminosity and the next-to-leading-order (NLO) cross section, 157 pb, obtained using MCFM~\cite{ref:mcfm,ref:mcfmtt}.
The QCD multijet component is fixed by the {\sc Pythia}-based simulation. Because of the 
$\met>25$ GeV requirement, the QCD 
yield is small and its uncertainty does not substantially affect the comparison.
To allow for a better comparison of the shapes of the distributions, 
the \PW +jets normalization is adjusted so that the total event yields in data and simulation 
agree, resulting in an increase by ${\approx} 40\%$ with respect to the inclusive next-to-next-to-leading order (NNLO)
value, $\sigma(\PW(\ell\nu))=31.3$ nb (obtained using {\sc FEWZ}~\cite{ref:fewz} and summed over all three lepton flavors). A similar effect is observed in the CMS \cPqt\cPaqt\ cross section measurement in 
the single-lepton channel, as discussed in  Ref.~\cite{ref:TOP-10-002}. 
This scaling is applied only for these
illustrative plots and is not relevant to the procedures used to obtain the background
contributions for the actual measurement.

The overall shapes of these and many other distributions are in qualitative
agreement with the simulation.
We have examined the \pt  distributions of the four leading jets,
the invariant mass distribution of the three leading jets, the lepton isolation distributions, the number
of \cPqb-tagged jets, and the transverse mass of the lepton-$\met$ system, which corresponds to the \PW\  boson
in most SM background processes. 
We conclude that the cores of the observed distributions are dominated by \cPqt\cPaqt\ and \PW+jets events.

Because the signal region involves the extreme tails of these distributions, which are 
difficult to simulate, 
the background predictions are based on control samples in the data
rather than on simulated event samples. 
The following sections describe these methods, which further 
probe the detailed kinematic features of the event sample. 

\subsection{Background determination using \HT and $\bmet/\sqrt{\HT}$}

Two kinematic variables that discriminate between SM backgrounds and 
new physics models such as SUSY, are \met and \HT. Using a rescaled version of 
$\met$ to minimize the correlation with \HT,
\begin{eqnarray}
\sMET \equiv \met/\sqrt{\HT}, 
\end{eqnarray}
where \HT is given by Eq.~(\ref{eq:HTdef}),
we can construct a set of control regions in the two-dimensional kinematic space 
of $(\HT, \sMET)$ and use them
to obtain a background estimate in the signal region.
The quantity \sMET\ can be interpreted as an approximate
\met significance in that the denominator is proportional to the uncertainty
on \met arising from jet mismeasurements \cite{PAS-JME-10-009}. 

The critical feature of these variables is that, due to the 
lack of significant correlation in the kinematic regions and event
samples used in this
analysis, their joint probability distribution
is, to a good approximation, simply the product
of the individual, one-dimensional distributions. 
These variables and a similar procedure were also used in the CMS 
opposite-sign dilepton SUSY search~\cite{ref:SUS-10-007}.

\begin{table}[tb!]
\caption{Definitions of the regions $A$, $B$, $C$, and $D$ with the loose and tight regions for the background
  estimation method using \HT and \sMET.\label{tab:ABCD-Regions}}
\begin{center}
\begin{tabular}{c|cc|cc}\hline\hline
 & \multicolumn{2}{c|}{Loose selection} & \multicolumn{2}{c}{Tight selection}\\
Region & \HT (GeV) & \sMET ($\sqrt{\rm GeV}$) & \HT (GeV) & \sMET ($\sqrt{\rm GeV}$)\\\hline
$A$    & $300<\HT<350$                 & $2.5<\sMET<4.5$           & $300<\HT<650$             & $2.5<\sMET<5.5$\\
$B$    & $\phantom{300<}\HT>400$      & $2.5<\sMET<4.5$           & $\phantom{300<}\HT>650$   & $2.5<\sMET<5.5$\\
$C$    & $300<\HT<350$                 & $\phantom{2.5<}\sMET>4.5$ & $300<\HT<650$             & $\phantom{2.5<}\sMET>5.5$    \\
$D$    & $\phantom{300<}\HT>400$       & $\phantom{2.5<}\sMET>4.5$ & $\phantom{300<}\HT>650$   & $\phantom{2.5<}\sMET>5.5$    \\
\hline\hline
\end{tabular}
\end{center}
\end{table}

\begin{figure}[tb!]
\begin{center}
\includegraphics[angle=0,width=0.35\textwidth]{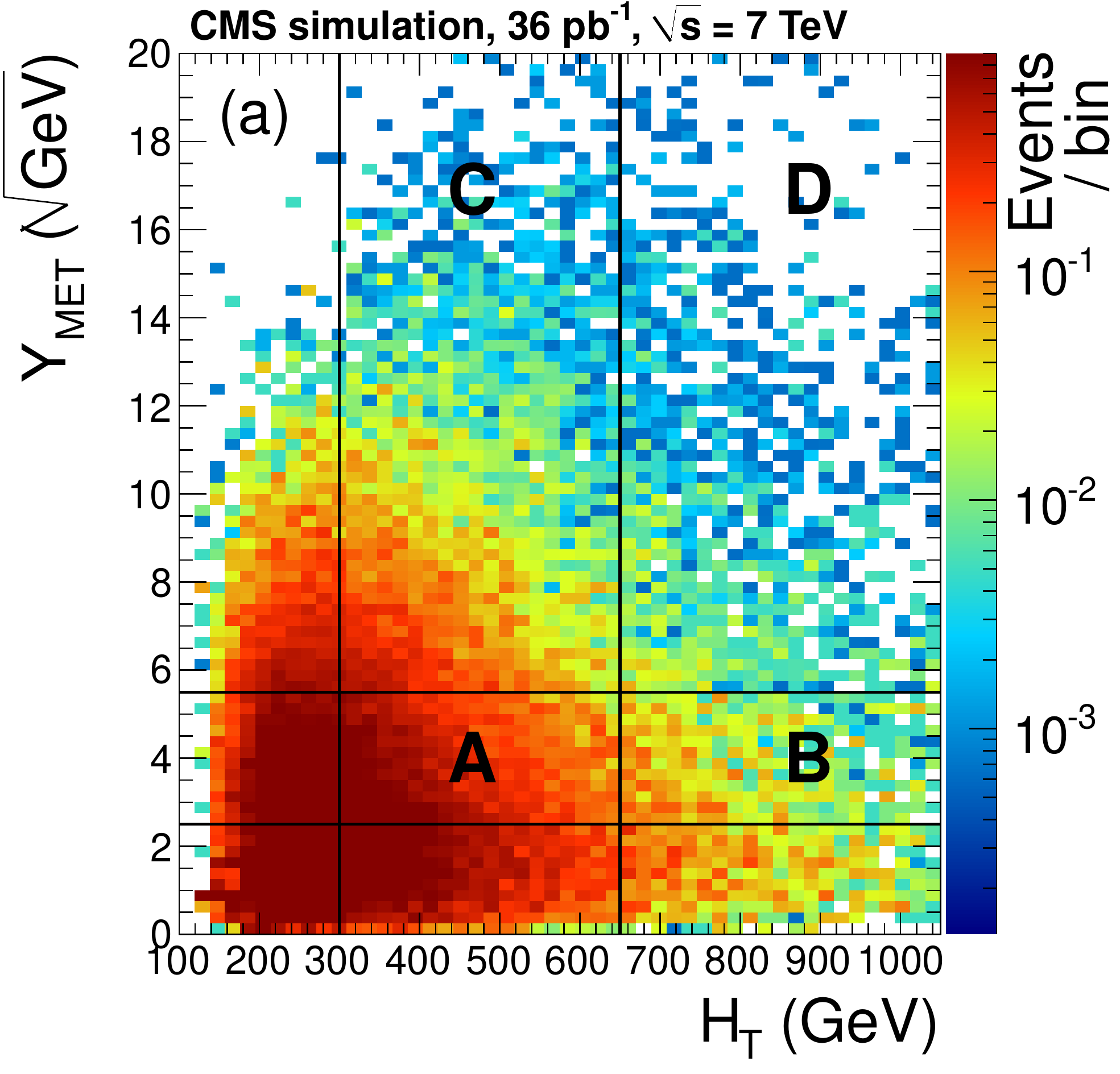}
\hspace{0.10\textwidth}
\includegraphics[angle=0,width=0.35\textwidth]{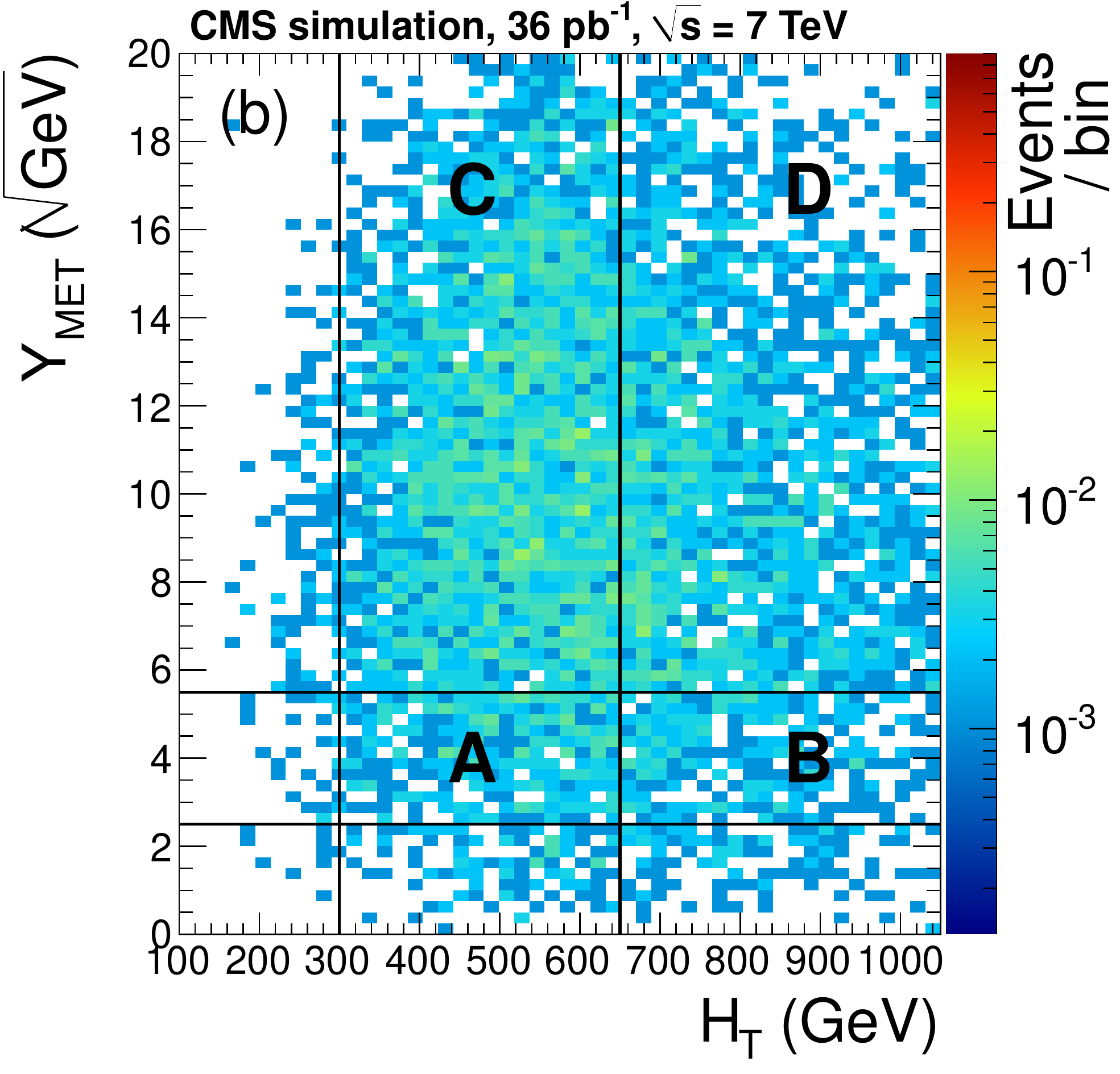} \\
\includegraphics[angle=0,width=0.35\textwidth]{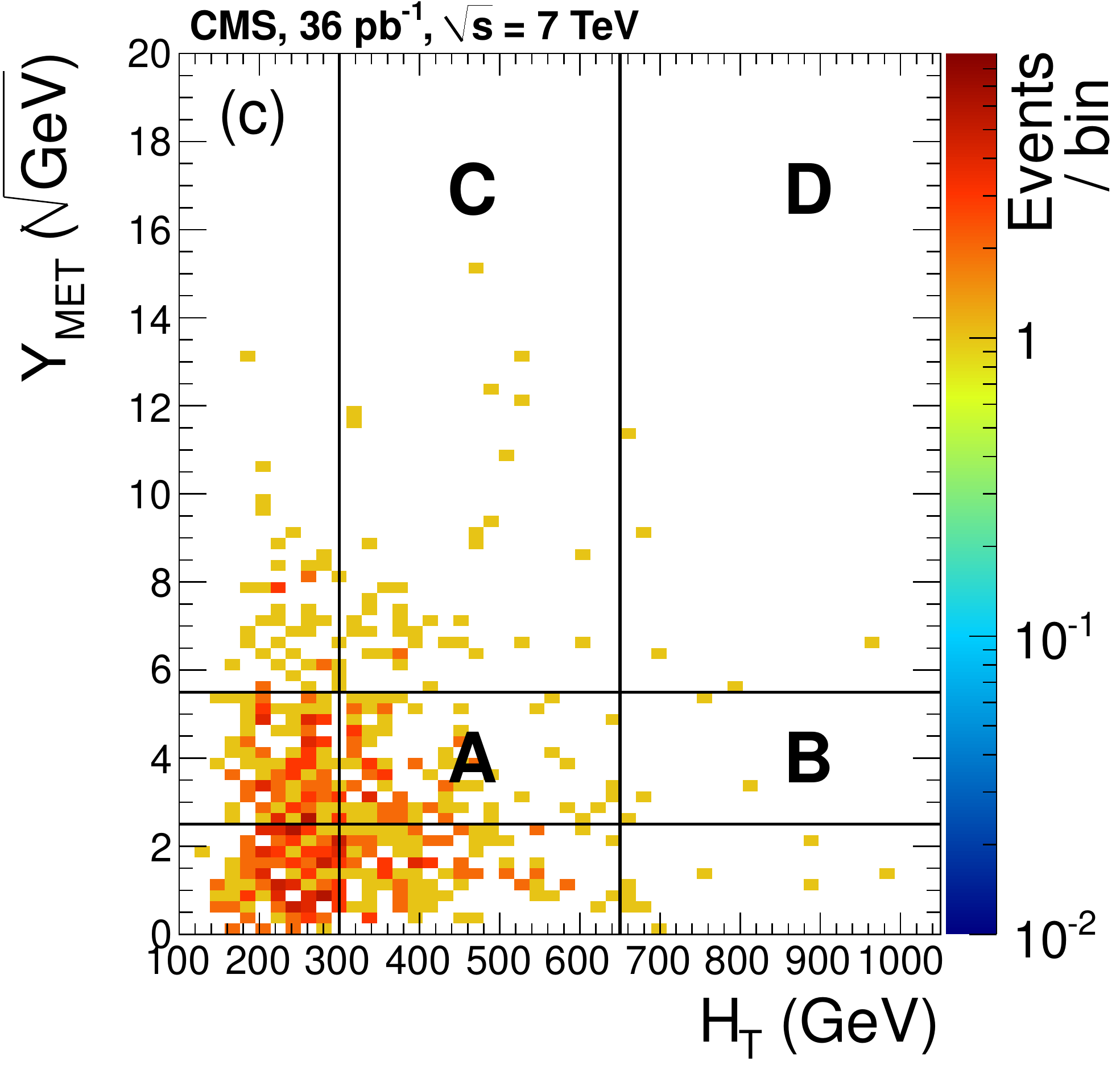}
\hspace{0.10\textwidth}
\includegraphics[angle=0,width=0.35\textwidth]{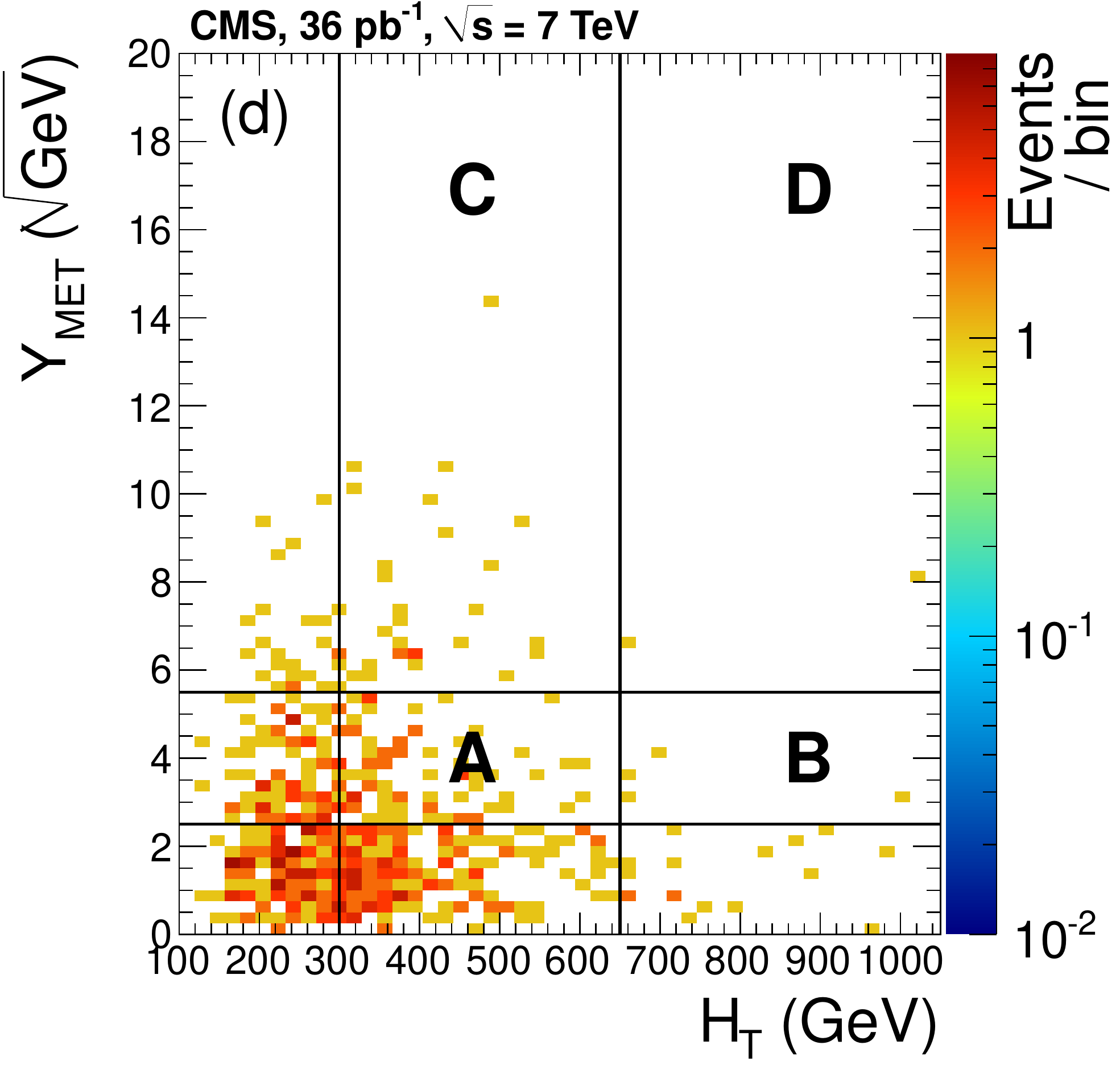}
\end{center}
\caption{Distributions of \HT vs.~\sMET\ for (a) the simulated total SM background (muon channel), (b) SUSY LM1 (muon channel),
(c) data in the muon channel, and (d) data in the electron channel. The control regions $ABC$ and the signal 
region $D$ are shown for the tight selection.}\label{fig:ABCD_2Dplots_mu}
\end{figure}

\begin{figure}[tb!]
\begin{center}
\includegraphics[angle=0,width=0.35\textwidth]{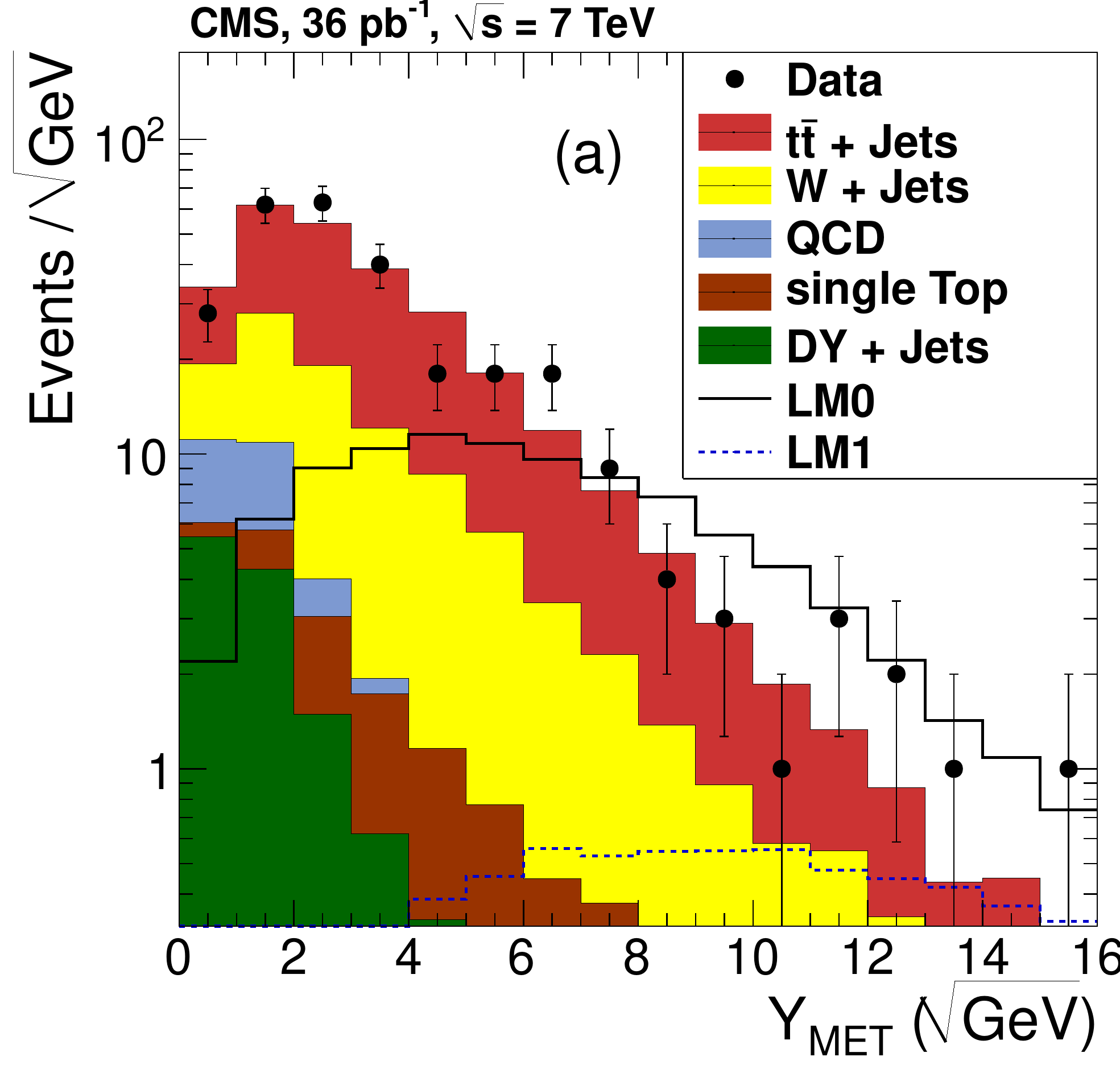}
\hspace{0.10\textwidth}
\includegraphics[angle=0,width=0.35\textwidth]{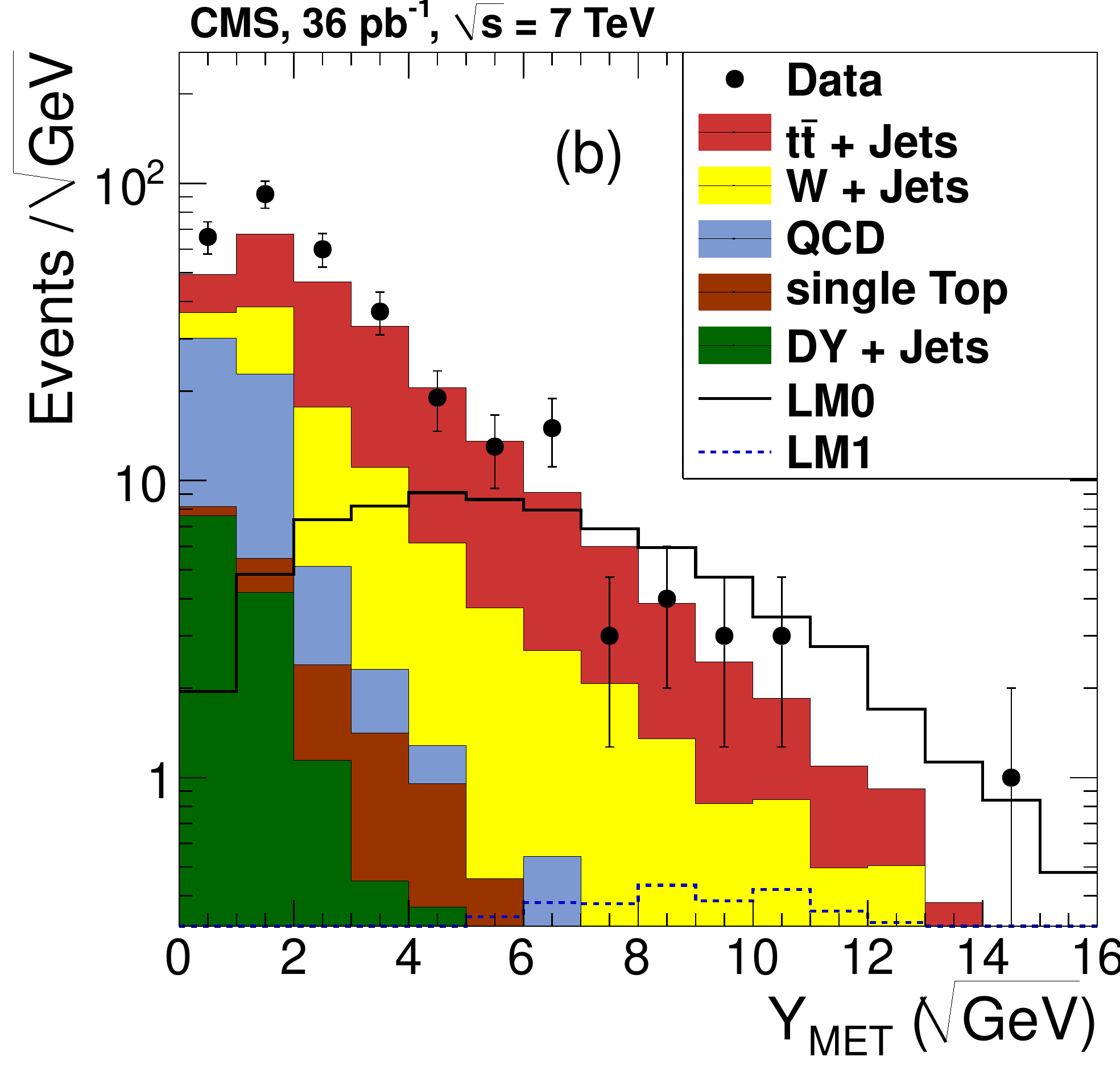}
\end{center}
\caption{Data and simulated (MC) distributions of $\sMET$ for (a) the muon channel and (b) the electron channel after the preselection  
and $\HT > 300$ GeV requirements. The simulated SM distributions are stacked; 
the distributions for the SUSY LM0 and LM1 benchmark models are overlaid.}\label{fig:ABCD_1Dplots}
\end{figure}

Table~\ref{tab:ABCD-Regions} defines ``loose'' and ``tight'' kinematic regions in the space
of $(\HT, \sMET)$, with four sub-regions denoted by $A$, $B$, $C$, and $D$ in each case.
Regions $A$, $B$, and $C$ have either low \sMET\ or \HT or both, while region
$D$, the signal region, has high values of both variables. 
Due to the very small correlation, the ratio of high-to-low \sMET\ 
events is nearly independent of \HT, and the number of SM background events
in region $D$ can be estimated from $N(D)_{\rm pred}=[N(C)/N(A)]N(B)$, 
where $N(i)$ denotes the number of events in region $i$. 
The tight selection was designed for SUSY models with small cross sections and
higher masses, while the loose selection is sensitive to large cross sections
and provides an additional handle for the comparison with the SM background
prediction.

Figure~\ref{fig:ABCD_2Dplots_mu} shows the two-dimensional distributions of 
\HT vs.~\sMET\ for the simulated signal and background samples, 
as well as for the data. 
For the signal samples, we have used the LM1 SUSY benchmark
model, normalized to the integrated luminosity of the data sample
using the NLO cross section given in Section~\ref{sec:EventSelection}. 
Events from this model are distributed at significantly higher values of \HT and
\sMET\ than those for the \cPqt\cPaqt\ and \PW +jets backgrounds, reflecting
the higher mass scales of the particles produced
in SUSY events.

The \sMET\ distributions for events satisfying the preselection requirements and $\HT>300$ GeV
(the lower \HT requirement on regions $A$ and $B$) are shown in 
Fig.~\ref{fig:ABCD_1Dplots} for the muon and electron channels. 
The shape of the \sMET\ distribution observed in the data agrees well with that predicted from the combined simulated event samples.
In the electron channel an excess is observed in the low \sMET\ bins --- a region 
where simulation predicts a contribution of QCD multijet events to the total background of about 
one third (compared to Fig.~\ref{fig:EventProperties} no explicit \met\ cut is applied in this selection).
The number of QCD multijet events in each of the four $ABCD$ regions, however, 
was measured from data and found to be small.

Table~\ref{tab:Yields_ABCDMethod} shows the event yields in the data 
and in the simulated background samples in the 
four regions, together with the predicted background level based on
this method. In both lepton channels, and for both the loose and tight selections, 
the predictions are statistically consistent with the observed yields in
the signal region.  Section~\ref{sec:SysErr} discusses the sources of systematic 
uncertainty, including small correlation effects and a potential bias 
from the small QCD multijet background.

\begin{table}[bt!]
\begin{center}
\caption{Data and simulated (MC) event samples: predicted and observed yields for the \HT vs.~\sMET\ background estimation method.
Tests of the method based on simulated event samples are included for comparison, but the actual background prediction is based on applying the
procedure to the data. The predicted yields, $N(D)_{\rm pred}$, in the signal region computed using the yields observed in
regions $A$, $B$, and $C$, as described in the text; these predictions are consistent with the observed yields in region $D$.  
The uncertainties shown are statistical only for simulation and statistical and systematic for the prediction in data.
The systematic uncertainties on $N(D)_{\rm pred}$ for data are discussed in Section~\ref{sec:SysErr}.\label{tab:Yields_ABCDMethod}}
\begin{tabular}{lccccc}\hline\hline
sample                         &   $N(A)$ &    $N(B)$ &   $N(C)$&  $N(D)$   & $N(D)_{\rm pred}$ \\ \hline
                     & \multicolumn{5}{c}{Loose selection} \\ \hline
$\mu$ channel: total SM MC     & $   25.1 \pm   0.6 $ & $  37.1 \pm   0.7 $ & $   19.3 \pm   0.5 $ & $  30.6 \pm   0.6 $ & $  28.5 \pm   1.1 $\\ 
$\mu$ channel: data            & $   30$            & $     35$             & $   25$              & $    30 $           & $  29.2 \pm   9.3\pm4.1$ \\ \hline
\Pe\  channel: total SM MC       & $   20.0 \pm   0.5 $ & $  31.5 \pm   0.9 $ & $   14.6 \pm   0.5 $ & $  23.6 \pm   0.5 $ & $  22.9 \pm   1.2 $\\ 
\Pe\  channel: data              & $   19$              & $  33$              & $   19$              & $   17$             & $  33.0\pm12.2\pm5.1$\\ \hline
                     & \multicolumn{5}{c}{Tight selection} \\ \hline
$\mu$ channel: total SM MC     & $   93.1 \pm   1.1 $ & $   8.7 \pm   0.4 $ & $  37.6 \pm   0.7 $ & $   3.4 \pm   0.2 $ & $    3.5 \pm   0.2 $\\ 
$\mu$ channel: data            & $   98$              & $   4$              & $  41 $             & $    5$             & $   1.7 \pm   0.9 \pm0.3$\\ \hline 
\Pe\  channel: total SM MC       & $   76.8 \pm   1.5 $ & $   6.5 \pm   0.3 $ & $  29.5 \pm   0.7 $ & $   2.9 \pm   0.2 $ & $   2.5 \pm   0.1 $\\ 
\Pe\  channel: data            & $    80$ & $    4$ & $   30$ & $    2$ & $    1.5\pm0.8\pm 0.3$\\ 
\hline\hline
\end{tabular}
\end{center}
\end{table}

In summary, we observe $N_{\rm obs}(\mu)=5$ and $N_{\rm obs}(\Pe)=2$ events in the signal region ($D$) 
after applying the tight selection requirements in the muon and electron channels. 
The predicted yields, based on the control regions in the \HT vs.~\sMET\ kinematic space, are
$N(\mu)=1.7\pm 0.9\ ({\rm stat.}) \pm  0.3\ ({\rm syst.})$ and  
$N(\Pe)=1.5\pm  0.8\ ({\rm stat.}) \pm  0.3\ ({\rm syst.})$,
which are statistically compatible with the observed yields. 
For the tight selection the predicted yields for the SUSY benchmark models described in 
Section~\ref{sec:EventSelection} are $22.5\pm0.7\ (\rm stat.)$ events for LM0
and $4.6\pm0.1\ (\rm stat.)$ events for LM1, with the yields divided approximately
equally between the muon and electron channels. 

\subsection{Background determination using the $\bmet$ and lepton \pt  distributions}
\label{ssec:BKG_LeptonSpectrum}

This section describes the lepton-spectrum method~\cite{ref:Pavlunin} for 
determining the shape of the $\met$ distribution  
from \cPqt\cPaqt\ and \PW+jets backgrounds with a single 
isolated lepton. These processes account for about 70\% of the total SM
contribution to the signal region, 
once the final selection requirements are applied. 
We also describe methods using control samples in data to 
measure most of the remaining background components. These 
arise mainly from (1) the feed-down of \cPqt\cPaqt\ dilepton events
(${\approx} 15$\%) and (2) either \cPqt\cPaqt\ or \PW+jets events with 
$\tau\to(\mu,\Pe)$ decays (${\approx} 15$\%). 
Although the background from QCD multijet events is very small,
we nevertheless measure this component using control samples in the data, 
because the uncertainties on the simulated QCD event samples are difficult 
to quantify. We rely on simulated event samples for the determination
of two backgrounds, single-top production and \cPZ+jets, whose
contributions are estimated to be below one event in total.

Two signal regions, loose and tight, are defined. 
The loose selection consists of the preselection
requirements (with, however, $\pt(\ell)>20$~GeV for both \Pe\  and $\mu$
for consistency), 
together with the requirement $\met>150$~GeV. For the tight selection,
we require $\met>250$~GeV and $\HT>500$~GeV, in addition to the
loose selection cuts. The tight selection is motivated by the fact that
for models with higher mass scales, the $\met$ and \HT distributions are shifted
upward, but the production cross sections fall rapidly. 

We distinguish between two forms of $\met$, genuine and artificial,
that contribute to the reconstructed $\met$ distribution. With 
the selection requirements used here, the dominant source
of $\met$ arises from the high-momentum 
neutrinos produced in \PW\  decay and
hence is genuine. Artificial $\met$ from jet mismeasurement is 
a much smaller effect but is not negligible and must 
be taken into account.
 
\begin{figure}[tbp!]
\begin{center}
{\includegraphics[angle=0,height=0.40\textwidth]{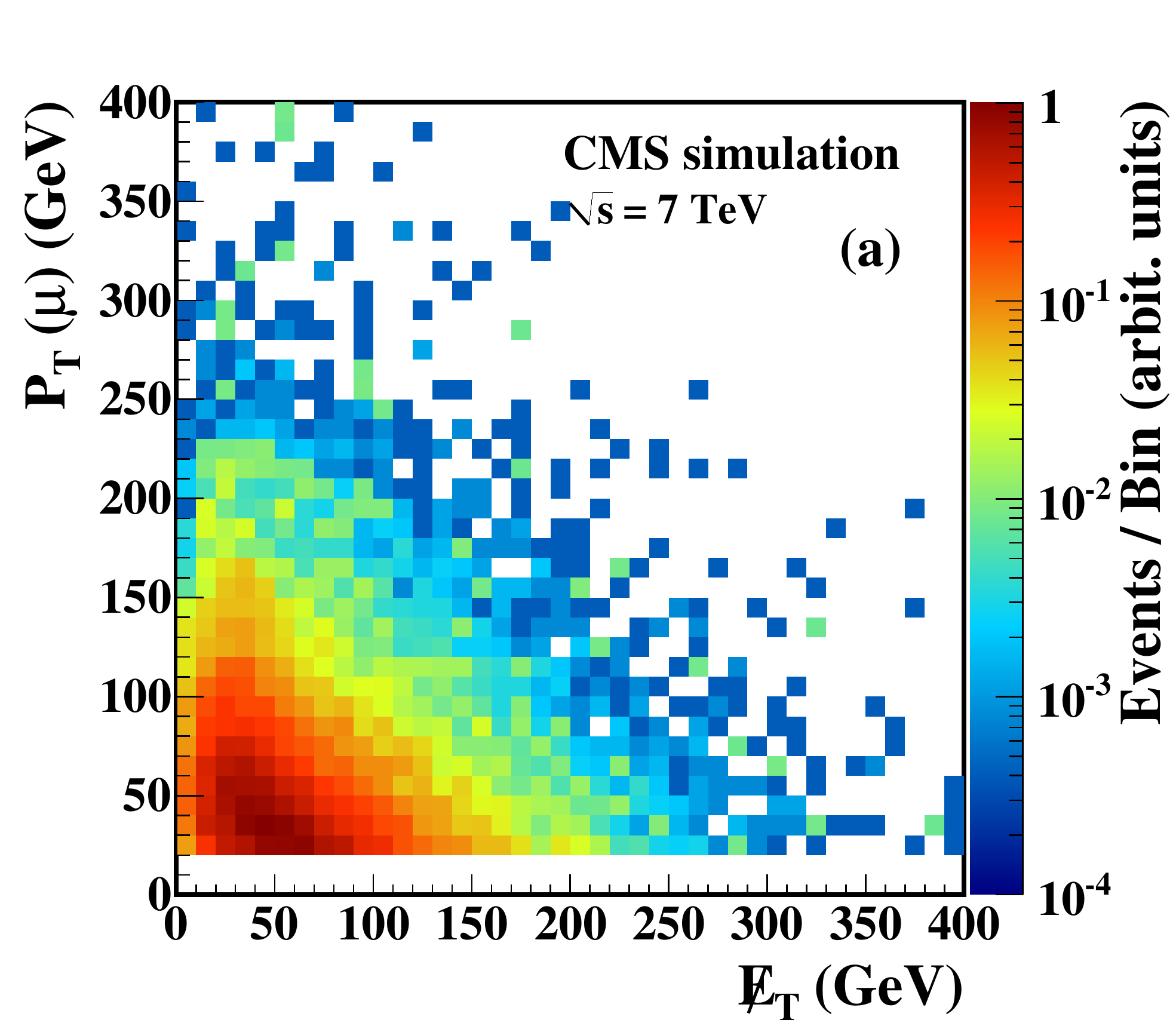} }
{\includegraphics[angle=0,height=0.40\textwidth]{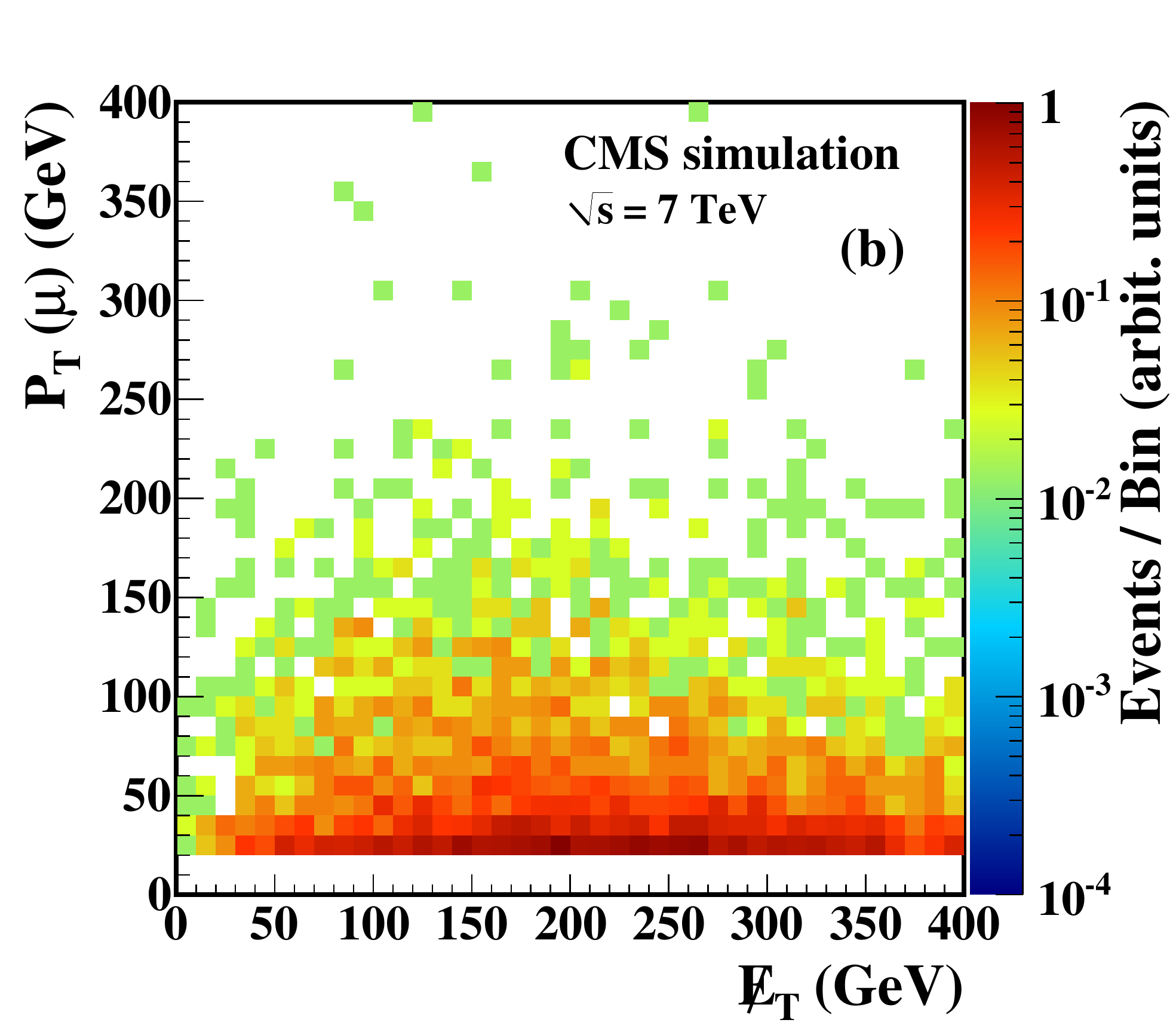} }
\end{center}
\caption{Distributions of muon \pt  vs.~$\met$ in the $\mu$ channel for (a) 
simulated \cPqt\cPaqt\ and \PW+jets events and (b) the LM1 SUSY benchmark model. In \cPqt\cPaqt\ and 
\PW+jets events,
the lepton \pt  and $\met$ in a given event are anticorrelated, but their distributions are 
very similar overall. In LM1, which is typical of many SUSY models,
the $\met$ distribution is much harder than the lepton spectrum, since it is dominated
by the production of two LSPs.}\label{fig:2D_pTvsMET}
\end{figure}

The physical foundation of the lepton-spectrum method is that,
when the lepton and neutrino are produced together in two-body
\PW\  decay  (either in \cPqt\cPaqt\ or in \PW+jets events) the
lepton spectrum is directly related to the $\met$ spectrum. 
With suitable corrections, discussed below, the lepton spectrum 
can therefore be used to predict the $\met$ spectrum.
In contrast, the $\met$ distribution in most SUSY models is dominated by 
the presence of two LSPs. The $\met$ distribution for such models 
extends to far higher values than the lepton spectrum. 
These points are illustrated in Fig.~\ref{fig:2D_pTvsMET}, which 
shows the relationship between lepton-\pt  and $\met$ distributions
in the laboratory frame for two simulated event samples: 
(a) the predicted SM mixture of \cPqt\cPaqt\ and \PW+jets events and (b) 
the SUSY LM1 benchmark model. 
As we will demonstrate, the lepton-spectrum method
provides a robust background prediction in the high $\met$ region, 
even in the presence of a large SUSY signal, because leptons
in SUSY events typically have much lower momenta than the LSPs.

To use the lepton spectrum to predict the $\met$ spectrum in single-lepton
SM background processes, three issues must be addressed: (1) the effect of 
\PW-boson polarization in both \cPqt\cPaqt\ and \PW+jets events, (2) the effect 
of the applied lepton \pt threshold, and (3) the effect of the difference between the 
experimental resolutions on the measurements of lepton \pt  and $\met$. 
We consider the polarization issues first. In each of these background processes, 
the leptons are produced in two-body \PW-boson decays, so that the 
momenta of the lepton and the neutrino are equal and opposite in the 
\PW\  rest frame, with angular distributions governed by the \PW\  polarization.
On an event-by-event basis, these momenta undergo identically 
the same sequence of Lorentz transformations from the \PW\  rest frame to the lab frame,
so in this sense the lepton spectrum automatically incorporates the effects of
the \cPqt-quark and \PW-boson \pt  distributions.
While the lepton and neutrino momenta are anticorrelated in the
laboratory frame on an event-by-event basis, 
the integrated distributions are very similar. 
If the angular distributions in the \PW\  rest frame were
forward-backward symmetric, the lab-frame $\met$ and lepton-\pt 
distributions would be identical. However, the helicity $\pm 1$ polarization
states of the \PW\  boson produce forward-backward asymmetries 
that can shift the lepton-\pt  spectrum with respect to the $\met$ spectrum.
 
We first consider \cPqt\cPaqt\ production, which is the largest background. In the decay of a top quark, 
\cPqt $\to$ \cPqb \PWp, the angular distribution of the (positively) charged lepton in the \PWp\  
rest frame can be written 
\begin{eqnarray}
\frac{dN}{d\cos\theta^*_{\ell}} = f_{+1}\frac{3}{8}(1+\cos\theta^*_{\ell})^2 +
f_{-1}\frac{3}{8}(1-\cos\theta^*_{\ell})^2+f_{0}\frac{3}{4}\sin^2\theta^*_{\ell},
\label{Eq:Wpolarization}
\end{eqnarray}
where $f_{+1}$, $f_{-1}$, and $f_{0}$ denote the polarization fractions associated with 
the \PW-boson helicities $+1$, $-1$, and 0, respectively. The angle $\theta^*_{\ell}$
is the polar angle of the charged lepton in the \PWp\  rest frame, measured 
with respect to a $z$-axis that is collinear with the momentum direction of the \PWp\  
in the top-quark rest frame. (In this expression, the azimuthal angle has been integrated
over, removing the interference terms between different helicity amplitudes.) 
The polarization fractions thus determine the angular distribution of 
the lepton in the \PW\  rest frame and, together with the Lorentz boosts, 
control the \pt distributions of the lepton and the neutrino in the laboratory
frame. 

The \PW\  polarization fractions in top-quark decay have been calculated~\cite{ref:Czarnecki}
with QCD corrections to NNLO, and the polarization is predominantly longitudinal.
For \cPqt $\to$ \cPqb \PWp\ these fractions are 
$f_{0}=0.687\pm0.005$, $f_{-1}=0.311\pm0.005$, 
and $f_{+1}=0.0017\pm0.0001$. The very small value of $f_{+1}$
is explained by the fact that, since $m_\cPqb/m_\PW\ll 1$, the \cPqb\ quark is highly relativistic 
and is in a nearly pure helicity $\lambda=-1/2$ state. Conservation of angular momentum along the
\cPqt-quark decay axis then forbids $\lambda = +1$ for the \PW\  boson. 
For the case of \cPaqt\ rather than \cPqt\ decay, the 
\PW-boson helicity fractions are swapped
between the $\lambda=\pm 1$ states, but 
the actual angular distribution of the lepton
is identical, because of the corresponding reversal of the
helicity state for the outgoing lepton, which has opposite charge.
These precise calculations reduce the uncertainties
associated with the \PW\  polarization in \cPqt\cPaqt\ events to a low level. 
The theoretical values are consistent with measurements from CDF, which 
obtained~\cite{ref:CDF_topDecayWpol} $f_{0}=0.88\pm 0.11\pm 0.06$ and 
$f_{+1}=-0.15\pm 0.07\pm 0.06$, expressed for the \PWp\  polarizations. 

If the \PW\  polarization were entirely longitudinal ($\lambda=0$), the angular 
distribution in the \PW\  rest frame would be forward-backward symmetric, and
the momentum spectra of the lepton and neutrino would be identical in the laboratory
frame. The effect of the substantial $\lambda=-1$ helicity component in \PWp\  decay is to 
give the lepton a preferred direction that is opposite to  
the \PWp\  momentum direction in the \cPqt\ rest frame. The
asymmetry produces a somewhat softer lepton spectrum than the \met spectrum, 
for both \cPqt\ and \cPaqt\ decays. The lepton spectrum therefore 
slightly underpredicts the high-$\met$ tail in \cPqt\cPaqt\ events, but the
effect is well understood and is included as a correction.

The \PW\  polarization in \PW +jets events exhibits a more complex behavior than 
that in \cPqt\cPaqt\ production. CMS has reported first measurements of these 
effects~\cite{ref:PAS-EWK-10-014}, which are consistent with
\ALPGEN and \MADGRAPH simulations predicting that 
the \PWp\  and \PWm\ bosons are both
predominantly left-handed in \PW +jets events at high \pt .
An NLO QCD calculation~\cite{ref:Blackhat} has demonstrated that 
the predicted polarization fractions are stable with respect to QCD
corrections. In contrast to \cPqt\cPaqt\ events, where only two of the \PW\  polarization states
are effectively present, all three \PW\  polarization states have significant
amplitudes in \PW +jets events. In addition, {\it both} of the \PWp\  and \PWm decay
polarization fractions for $\lambda = -1$ are in the range 55--70\% and increase
gradually with $\pt(\PW)$. Because the \PW$^\pm$ daughter leptons have opposite helicities,
this leads to {\it opposite} asymmetries for the lepton angular 
distributions. The cancellation in the asymmetries is not perfect, however, mainly 
because the \PWp\  cross section in \Pp\Pp\ collisions is substantially higher than that for \PWm
production. With the $\met$ and lepton \pt  requirements applied in the 
analysis, the relevant \PW\  bosons have $\pt(\PW)>150$ GeV. The systematic uncertainties 
associated with these effects are discussed in Section~\ref{sec:SysErr}.

The relationship between the lepton \pt  spectrum and the $\met$ distribution
is also affected by the threshold ($\pt>20$ GeV) applied to the leptons. 
Because of the anticorrelation between the lepton \pt  and the $\met$, the threshold
requirement removes SM background events at high $\met$ but not the events with
high-\pt  leptons that are used to predict this part of the $\met$ spectrum. 
For the \cPqt\cPaqt\ background, this effect
partially compensates for the bias from the \PW\  polarization. For \PW +jets events, 
in contrast, the polarization effects for \PWp\  and \PWm\ approximately cancel, but the
lepton \pt  threshold shifts the predicted yield upward.
The key point is that the effects of both the polarization and the lepton \pt  
threshold can be reliably determined.

Finally, the resolution on the reconstructed $\met$ is poorer than that for the lepton \pt ,
so the $\met$ spectrum will be somewhat broadened with respect to the 
prediction from the lepton spectrum. 
We measure $\met$ resolution functions (templates)
in the data using QCD multijet events, and use them to smear the measured lepton momenta.
The templates are created for events with $\ge 4$ jets
and are characterized by the \HT range of the events. Because the
templates are taken from data, they include not only the intrinsic
detector resolutions, but also the effects of cracks and acceptance.
The overall effect of the smearing is modest, changing the background prediction by 
5--15\%, depending on the $\met$ threshold applied. 

The background predictions based on control samples in data require
correction factors to account for a specific set of effects. For
the single-lepton backgrounds, the effects of the \PW\  polarization, 
the lepton \pt  threshold for the signal region,
and the $\met$ energy scale are to produce understood shifts in the 
$\met$ spectrum relative to the lepton spectrum.
The correction factors also account for a small contamination 
of the single-lepton control sample from dilepton and single-$\tau$ events
with high \pt  leptons, ${\approx} 2\%$
for the tight selection. Overall, the lepton \pt  spectra from these processes
are much softer than the corresponding $\met$ distributions, and the
background predictions must be obtained from separate control samples.

To account for these effects, the raw predicted yields are  multiplied 
by correction factors, obtained from simulated event samples. The
non-single-lepton backgrounds themselves are estimated from independent
methods described below.
For the tight selection the correction factors for the single-lepton
background are near unity: $0.88\pm0.07$  for the muon channel 
and $0.89\pm0.08$ for the electron channel. In the loose selection, the factors 
are $0.62\pm0.02$ (muons) and $0.70\pm0.02$ (electrons).
The uncertainties on the correction factors quoted here 
are statistical only. Systematic uncertainies are discussed 
in the following section.

\begin{figure}[tb!]
\begin{center}
\subfigure{\includegraphics[angle=0,height=0.40\textwidth]{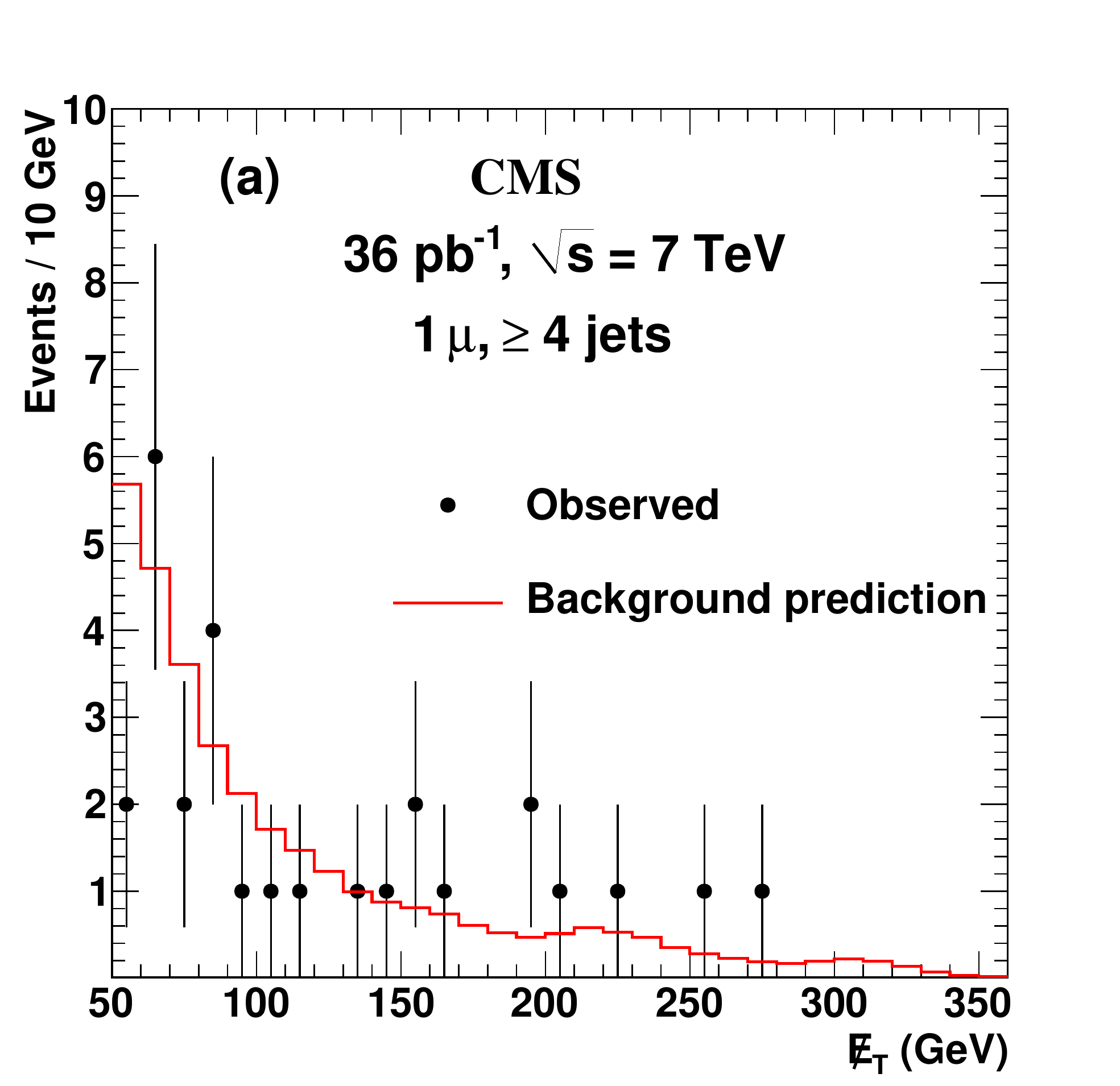} }
\subfigure{\includegraphics[angle=0,height=0.40\textwidth]{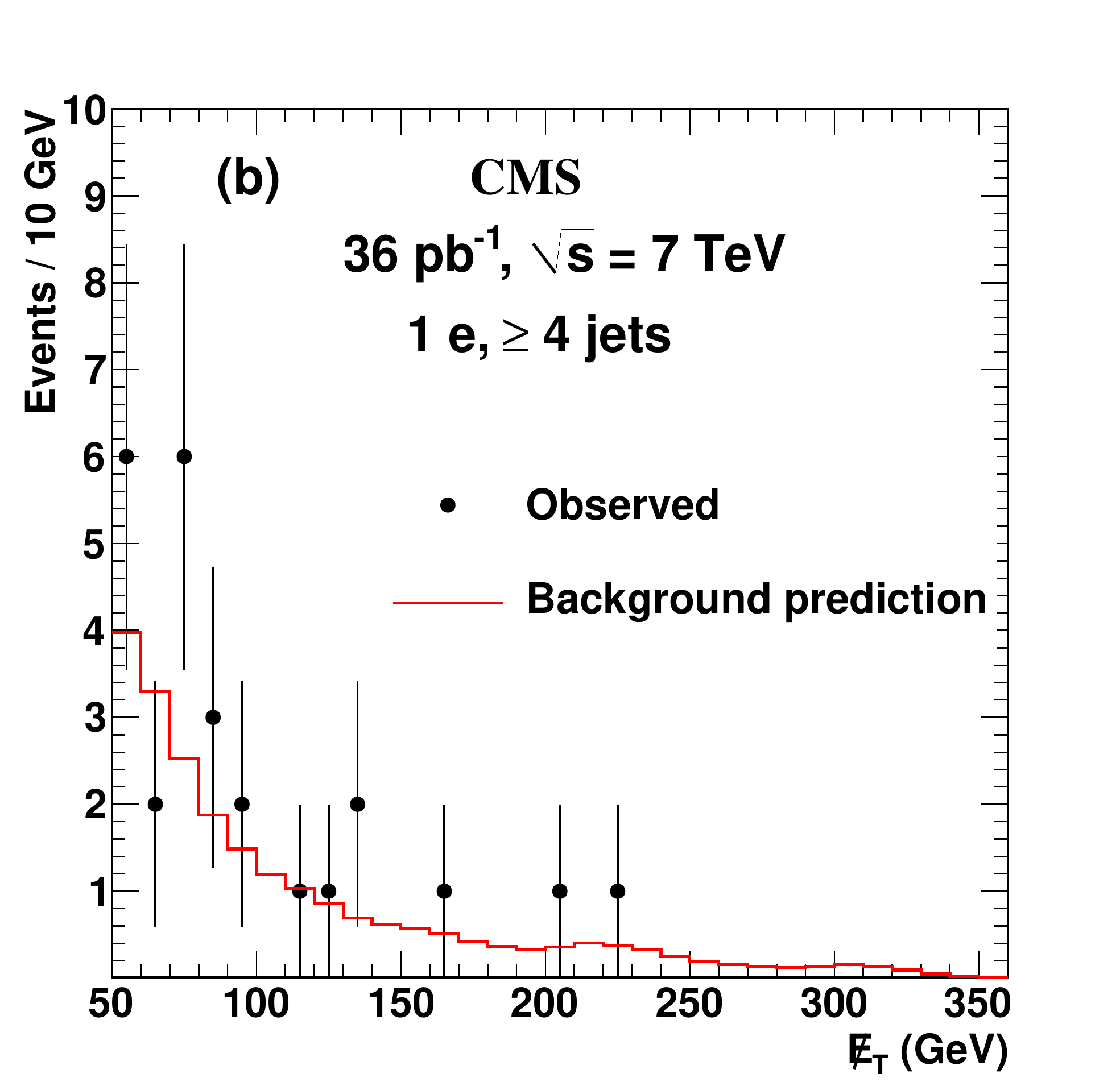} }
\end{center}
\caption{Measured vs.~predicted $\met$ distributions in (a) muon and (b) electron channels, with tight selections applied.
The data are shown as points with error bars, while the prediction from the resolutions-smeared lepton
spectrum is shown as the histogram. The predicted single-lepton SM background yield for $\met>250$ GeV 
is obtained from these curves, after applying a correction factor described in the text.}\label{fig:DataPrediction}
\end{figure}

Figure~\ref{fig:DataPrediction} shows the $\met$ distributions
for the data in the muon and electron channels, with all 
of the tight selection requirements applied, except that on $\met$ itself. 
The predicted $\met$ distribution is obtained by applying the $\met$-smearing 
procedure, as described above, to the raw single-muon 
\pt spectrum. The predicted single-lepton background is in good 
agreement with the observed $\met$ spectra. 
The background predictions shown in Fig.~\ref{fig:DataPrediction} do not 
include the smaller contributions from non-single-lepton
sources.

\begin{table}[tb!]
\begin{center}
\caption{Loose selection: predicted and observed yields in the signal region (pre-selection, $\met>150$ GeV). 
The quoted uncertainties are statistical and systematic. All background contributions
are determined from control samples in the data, except for the single-top and \cPZ +jets 
contributions, which are obtained from simulated event samples.\label{tab:Data_Loose}}
\begin{tabular}{lcc}\hline\hline
Sample                  &  $\ell=\mu$        &  $\ell=\Pe$            \\ \hline 
Predicted SM 1 $\ell$   &  $11.1\pm2.8\pm3.0$      &  $8.8\pm2.9\pm2.4$            \\
Predicted SM dilepton   &  $1.0\pm0.6\pm0.1$       &  $0.8\pm0.5\pm0.03$             \\          
Predicted single $\tau$ &  $2.1\pm0.6\pm0.2$       &  $2.2\pm0.5\pm0.3$             \\ 
Predicted QCD background &  $0.18\pm0.13\pm0.09$    &  $0.0^{+0.38}_{-0.0}\pm0.19$      \\ 
Predicted single top, \cPZ +jets &  $0.4\pm0.1\pm 0.2$    &  $0.4\pm0.1\pm0.2$      \\ \hline
Total predicted SM      &   $14.8\pm2.9\pm3.0$    &  $12.2\pm3.0\pm2.4$            \\
Observed signal region  &   $16$             &  $13$                    \\
\hline\hline
\end{tabular}
\end{center}
\end{table}

\begin{table}[tb!]
\begin{center}
\caption{Tight selection: predicted and observed yields in the signal region (pre-selection, $\met>250$ GeV, $\HT>500$ GeV).
The quoted uncertainties are statistical and systematic. All background contributions
are determined from control samples in the data, except for the single-top and \cPZ +jets 
contributions, which are obtained from simulated event samples.\label{tab:Data_Tight}}
\begin{tabular}{lcc}\hline\hline
Sample                  &  $\ell=\mu$        &  $\ell=\Pe$            \\ \hline 
Predicted SM 1 $\ell$   &       $1.5\pm1.1\pm0.7$  &   $1.1\pm0.8\pm0.5$        \\
Predicted SM dilepton   &       $0.0^{+0.3+0.23}_{-0.0-0.0}$ & $0.0^{+0.4+0.14}_{-0.0-0.0}$       \\          
Predicted single $\tau$ &       $0.16\pm0.10\pm0.20$     &  $0.27\pm0.20\pm0.20$   \\
Predicted QCD background &      $0.09^{+0.12}_{-0.09}\pm0.04$    & $0.0^{+0.16}_{-0.0}\pm0.08$       \\ 
Predicted single top, \cPZ +jets &      $0.05^{+0.05}_{-0.04}\pm0.05$    & $0.01\pm0.003\pm0.01$       \\ \hline
Total predicted SM      &       $1.8\pm1.1\pm0.8$     & $1.4\pm0.9\pm0.5$        \\
Observed signal region  &       $2$                &       $0$              \\
\hline\hline
\end{tabular}
\end{center}
\end{table}

Tables~\ref{tab:Data_Loose} and \ref{tab:Data_Tight} list the observed yields
and the predicted SM background contributions for the loose and tight 
selection requirements. The event yields observed in the data are largely accounted
for by the direct single-lepton backgrounds. As we have noted, however, the
lepton-spectrum method does not comprehensively predict all of the 
backgrounds to the single-lepton sample, and non-negligible backgrounds
arise from other sources, including several categories of 
dilepton events, events with $\tau\to\ell$ decays (either from \cPqt\cPaqt\ 
or \PW +jets events), and QCD multijet processes. These contributions
are also estimated using control samples in the data, as discussed below.
The background from single-top production and Drell-Yan/\cPZ  +jets is very small
for the loose selection and is negligible for the tight selection. These 
contributions are estimated from Monte Carlo samples, with systematic
uncertainties taken to be 50\% (100\%) for the loose (tight) selection. 
Because of their small absolute size, these uncertainties
have a negligible effect on the total background uncertainty. 

The dilepton background (including the $\tau$ as one of the leptons) 
can be divided into four contributions: 
(1) $2\ell$ with one ignored lepton,
(2) $2\ell$ with one lost lepton,
(3) $\ell+\tau$ with $\tau\to{\rm hadrons}$, and 
(4) $\ell+\tau$ with $\tau\to{\rm lepton}$. 
An ignored lepton is one that is reconstructed but fails either the
lepton identification requirements or the \pt  threshold requirement. 
A lost lepton is one that is either not reconstructed or is out of the
detector acceptance. Events from processes (1) and (3) account for most of the dilepton
background. All of the estimates of the dilepton feed-down backgrounds 
begin with control samples of reconstructed 
dilepton events in the $\Pe\Pe$, $\Pe\mu$, and $\mu\mu$ channels. 
The $\met$ distributions in these control samples in data, when suitably modified
to reflect the loss of a lepton or the presence of 
a leptonic or hadronic $\tau$ decay,
provide an accurate description of the shape of the $\met$
distribution of the background. Simulated event samples are used to determine, 
for the four processes described above, 
the ratio $r_i=N_{\rm feed}^i/N_{\rm control}$
of the number of events feeding down to the single-lepton channel to the number of events
observed in the control sample. This procedure effectively 
normalizes all such feed-down contributions to the control samples in data.
In all cases, care is required to ensure that the control sample
is not contaminated by QCD background. 
Estimates for the $\tau\to\ell$ single-lepton backgrounds from
\cPqt\cPaqt\ and \PW +jets processes are based on a similar 
procedure as that used for the dilepton backgrounds, but in this case 
the single-lepton sample itself is used as the control sample.

We define correction factors for the dilepton and single-$\tau$ background predictions.
In the loose selection, these corrections range from 0.86 to 0.94, with ${\approx} 10$\% uncertainty. 
For the tight selection, the correction factors are 
typically $\sim0.5$, with a large (${\approx} 75\%$) systematic uncertainty.
This correction has almost no effect on the final result, because the
background from these sources is small compared to the single-\Pe\  and 
single-$\mu$
backgrounds (see Tables~\ref{tab:Data_Loose} and \ref{tab:Data_Tight}).

Background from QCD multijet events is suppressed to a level well below one event
in both the loose and tight selections.
To estimate the QCD background,  
we use the two-dimensional distribution of $\met$ and the relative
lepton isolation, $I/\pt(\ell)$ (Section~\ref{sec:EventSelection}),
which are essentially uncorrelated. 
Using a QCD-dominated sample with $\met < 25$ GeV, we 
measure the ratio of the number of leptons passing the isolation requirement
($I/\pt(\ell) < 0.1$) to the number in an isolation sideband 
($0.2 < I/\pt(\ell) < 0.5$). Events that pass the $\met$ requirements for the signal region, 
but are in the isolation sideband, are then scaled by this measured ratio.

The precision of the QCD background prediction is 
limited by the small number of 
sideband leptons in the high-$\met$ region.
Two such muon events are found with $\met>150$ GeV, 
one of which also passes the $\met>250$ GeV requirement,
while there are no electron events. 
This procedure tends to overestimate the QCD background, because
events from electroweak and \cPqt\cPaqt\ processes can contaminate
the high-$\met$ isolation sideband and the isolated, low-$\met$ sample.
In addition, for the tight selection the 
measurement is performed using a loosened \HT requirement
of $120$~GeV for muons and $300$~GeV for electrons, since the isolation
sideband is sparsely populated. Despite these potential 
overestimates, the background predictions and their 
uncertainties, listed in Tables~\ref{tab:Data_Loose} and 
\ref{tab:Data_Tight}, are small, well below one event.

Although very few QCD background events contribute to the signal region
at high $\met$, such events can affect the control region used to estimate 
the single-lepton background from \cPqt\cPaqt\ and \PW +jets events.
That control sample is selected without a $\met$ requirement. 
In fact, requiring a minimum value of $\met$, say $\met>25$~GeV, 
would tend to remove events with high-\pt  leptons,
which are precisely those used to predict the high-$\met$ tail. 
The QCD contamination in the muon sample is very small, but there is
significant contamination from QCD in the electron sample at low $\met$.
We have therefore used only the \pt  spectrum from the muon control sample 
to predict the rates for both the electron and muon signal regions. 
The scaling from the muon to the electron samples is obtained by fitting their
ratio in the data over the range $60 < \met < 140$~GeV, with systematic uncertainties
evaluated by varying the fit range. The resulting correction factor, 
$N(\Pe)/N(\mu)=0.70\pm 0.15$, is consistent with the value obtained 
using simulated event samples. 

In summary, the background yields listed in Tables~\ref{tab:Data_Loose} and
\ref{tab:Data_Tight} are consistent with the total 
background predicted in each selection, for both the electron and muon channels.
In the loose selection, 16 muon events are observed in data compared with 
$14.8\pm2.9\pm3.0$ predicted, while 13 electron events are observed in data 
compared with $12.2\pm3.0\pm2.4$ predicted. In the tight selection, 
2 muon events are observed in data compared with $1.8\pm 1.1\pm0.8$ predicted, while
0 electron events are observed compared with $1.4\pm0.9\pm0.5$
predicted. The interpretation of these results in terms of SUSY 
models is discussed in Section~\ref{sec:Results}.

\section{Systematic Uncertainties}
\label{sec:SysErr}

This section discusses the systematic uncertainties associated with the \HT vs.~\sMET\ and the lepton-spectrum
methods. The uncertainties fall into two main categories: the uncertainties on the 
background estimates and the uncertainties on the overall acceptance and 
efficiency factors that are used to convert the observed yields into upper limits on SUSY cross sections. 

The \HT vs.~\sMET\ method predicts the background yield in the signal region ($D$) as a function of the yields in the
control regions $A$, $B$, and $C$. The systematic uncertainty arises from the possibility of a non-zero 
correlation between the kinematic variables and can be expressed in terms of 
small departures of the quantity $\kappa\equiv N(A)N(D)/N(B)N(C)$ from unity. 
Monte Carlo simulation predicts values of $\kappa$ close to unity for 
\cPqt\cPaqt, \PW+jets, and single-top production, as well as for the sum of all backgrounds.
As an additional check, this behavior of $\kappa$ has also been verified for the three-jet samples, 
which are not used in the analysis.

We have evaluated the effect on $\kappa$ from an extensive list of uncertainties.
Reconstruction-related uncertainties include the jet (and \met) energy scales, the jet-energy resolution, 
the amount of energy in the calorimeter not clustered into jets, the jet reconstruction efficiency,
the lepton-\pt\  scale, and the \pt  dependence of the efficiency.
Physics-related uncertainties under consideration were related to the background composition (\cPqt\cPaqt\ vs.~\PW+jets), to the amount of QCD background subtracted from each control region, and to the parton distribution functions.
The small deviation of the central value of $\kappa$ from unity 
predicted by the simulation has been added as an additional uncertainty.
These sources of systematic uncertainties are taken to be uncorrelated and the contributions are added in quadrature.
For the loose selection, the total systematic uncertainties affecting the background prediction in the muon and electron channels are $14\%$ and $16\%$, respectively. 
The corresponding numbers for the tight selection are $16\%$ ($\mu$) and $21\%$ (\Pe).

The systematic uncertainties on the lepton-spectrum background predictions are substantially larger,
and they increase from the loose to the tight selection. The dominant uncertainty
is associated with the jet and $\met$ energy scale~\cite{ref:PAS-JME-10-010}. If this scale shifts relative to the
lepton \pt  scale, the predicted number of events above the $\met$ threshold for the signal
region will change. The 5\% uncertainty on this scale propagates to a 22\% uncertainty
for the loose selection and a 37\% uncertainty for the tight selection. 

The precision with which the lepton spectrum prediction matches the $\met$ spectrum is determined
by a set of related effects, as described in Section~\ref{ssec:BKG_LeptonSpectrum}. 
The helicity fractions for \PW\  bosons produced in \cPqt\cPaqt\ events are predicted in 
the SM to high precision; when these uncertainties are propagated through the analysis,
the effect on the background prediction is negligible. As a test, we have varied
the polarization factors through a range that is about ten times the theoretical
uncertainties quoted in Ref.~\cite{ref:Czarnecki}. This leads to only a 2\%
effect for the loose selection and a 4\% effect for the tight selection. 
We have also varied the \cPqt-quark \pt  spectrum to study the effect of the boost on any
differences arising from the polarizations. This variation is constrained by
the agreement between data and simulation for the \PW-boson \pt  spectrum,
and leads to 5\% (loose) and 7\% (tight) uncertainties on the background yield. 
In addition, we vary the \cPqt\cPaqt\ cross section by $\pm 30\%$ and the 
\PW +jets cross section by $\pm50\%$ and measure the effect on the background prediction 
in simulated event samples (12\% for loose, 16\% for tight selection). 

To account for the \PW\  polarization uncertainties, we have chosen three
variations of the polarization fractions: (1) 100\% variation
on $f_{-1}-f_{+1}$, for both \PWp\  and \PWm together (this
is equivalent to an approximately 30\% variation of the individual
polarization fractions); (2) 10\% variation of $f_{-1}$ 
and $f_{+1}$, with constant sum, for the \PWp\  polarization,
holding the \PWm polarization fixed, and vice-versa; and 
(3) 100\% variation of the longitudinal polarization fraction,
$f_{0}$, for both \PWp\  and \PWm. Each variation
is applied in the same manner in three bins of $\pt(\PW)$:
50--100 GeV, 100--300 GeV, and $>300$ GeV. 
We do not vary the polarization of events with $\pt(\PW)<50$ GeV
since these have a negligible contribution to the 
selected event sample.  The sum of all three variations
in quadrature yields a 7\% systematic uncertainty for the
loose selection and a 14\% uncertainty for the tight selection. 

The systematic uncertainty arising from the possible incorrect modeling of the 
\pt  dependence of the lepton reconstruction and identification efficiency is estimated 
to be ${\approx} 4\%$. The effect of a potential mismodeling of jet reconstruction 
efficiencies is found to be negligible. The total systematic uncertainties on 
the lepton-spectrum method for predicting the
single-lepton background is 27\% for the loose selection and 44\% for the tight 
selection. These do not include the uncertainties on the separate 
estimates for the dilepton and $\tau$ backgrounds based on control samples. These additional predictions
are assigned a systematic uncertainty based on tests with simulated samples, including both the
statistical uncertainty due to finite simulated event samples and any observed shift with respect to the
true values. 

The effect of the $\met$ resolution 
(smearing) in simulated \cPqt\cPaqt\ events (using simulated QCD $\met$ templates) is to increase the background prediction 
by about 10\% for the loose selection. The smearing from the data has been seen to increase the background prediction 
slightly more, by about 15\%. We have increased the size of the template binning in \HT by factors of two and five and recomputed 
the resolution smearing in each case. The effects are negligible, demonstrating that the
prediction is insensitive to the details of the templates. 

To translate from the observed event yields to cross section limits, we must incorporate
the effects of the signal efficiencies and acceptance. These quantities are taken from the
simulated event samples, with cross-checks performed using the data as a validation.  
The uncertainties include those on 
the modeling in simulation of the lepton trigger and identification efficiencies (5\%), 
on the jet and $\met$ energy scales (17\% in the lepton-spectrum method), 
on the possible variation of parton density functions (negligible), 
and on the luminosity (4\%).
The total systematic uncertainty on the efficiency and acceptance is 20\%. 

%% file: Results_Final.tex
\section{Results and Interpretation}
\label{sec:Results}

Both of the methods used to determine the SM background  
predict yields that are compatible with the observed number of events.
In the absence of a signal, we proceed to set exclusion limits
on SUSY parameter space.  

The potential signal contamination of the control samples in the data is model dependent and must be assessed separately for each signal-model hypothesis. 
We have performed a scan over CMSSM model points 
and have determined the number of such
events that enter the control regions of our measurements. For 
the lepton-spectrum method, the contamination is small,
0.05 events on average. However, the method using control regions in the
\HT vs.~\sMET\ plane suffers a much greater contamination
of the control regions, especially for models with large cross 
sections. For the purpose of setting limits on the CMSSM, we have therefore
used the values obtained from the lepton-spectrum method. 

Combining the yields in the lepton-spectrum method from the \Pe\  and $\mu$ 
channels, we observe 29 events in the loose selection and 2 events 
in the tight selection. The predicted SM background is $27.0\pm 7.0$
events and $3.2\pm2.3$ events for the loose and tight selections, respectively.
(Because the muon spectrum is used as a control sample for obtaining
the single-lepton background in both the \Pe\  and $\mu$ channels,
as discussed in Section~\ref{sec:BackgroundDetermination}, the combined prediction
reflects the fact that the uncertainties between these two
channels are highly correlated.) Applying the Feldman-Cousins 
method~\cite{ref:FeldmanCousins}, which takes into account 
the number of events in the control samples using the profile likelihood 
ratio~\cite{ProfileLikelihood} to handle nuisance parameters, yields
a $95\%$ confidence level (CL) upper limit of 20.4 signal 
events (loose selection) and 3.8 signal events (tight selection).
The central value and $\pm 1~\sigma$ range of the expected limits 
are obtained by applying the same method to MC pseudo-experiments.
For comparison, the SUSY LM0 model predicts $64\pm 1$ events for the 
loose selection and $11.2\pm0.3$ events for the tight selection (\Pe\  and $\mu$
channels combined). The LM1 model, for which the yields are
$8.7\pm 0.1$ events (loose, \Pe +$\mu$) and 
$4.2\pm 0.1$ events (tight, \Pe +$\mu$), 
is at the edge of the sensitivity of the analysis.

\begin{figure}[tbp!]
\begin{center}
\subfigure{\includegraphics[angle=0,height=0.55\textwidth]{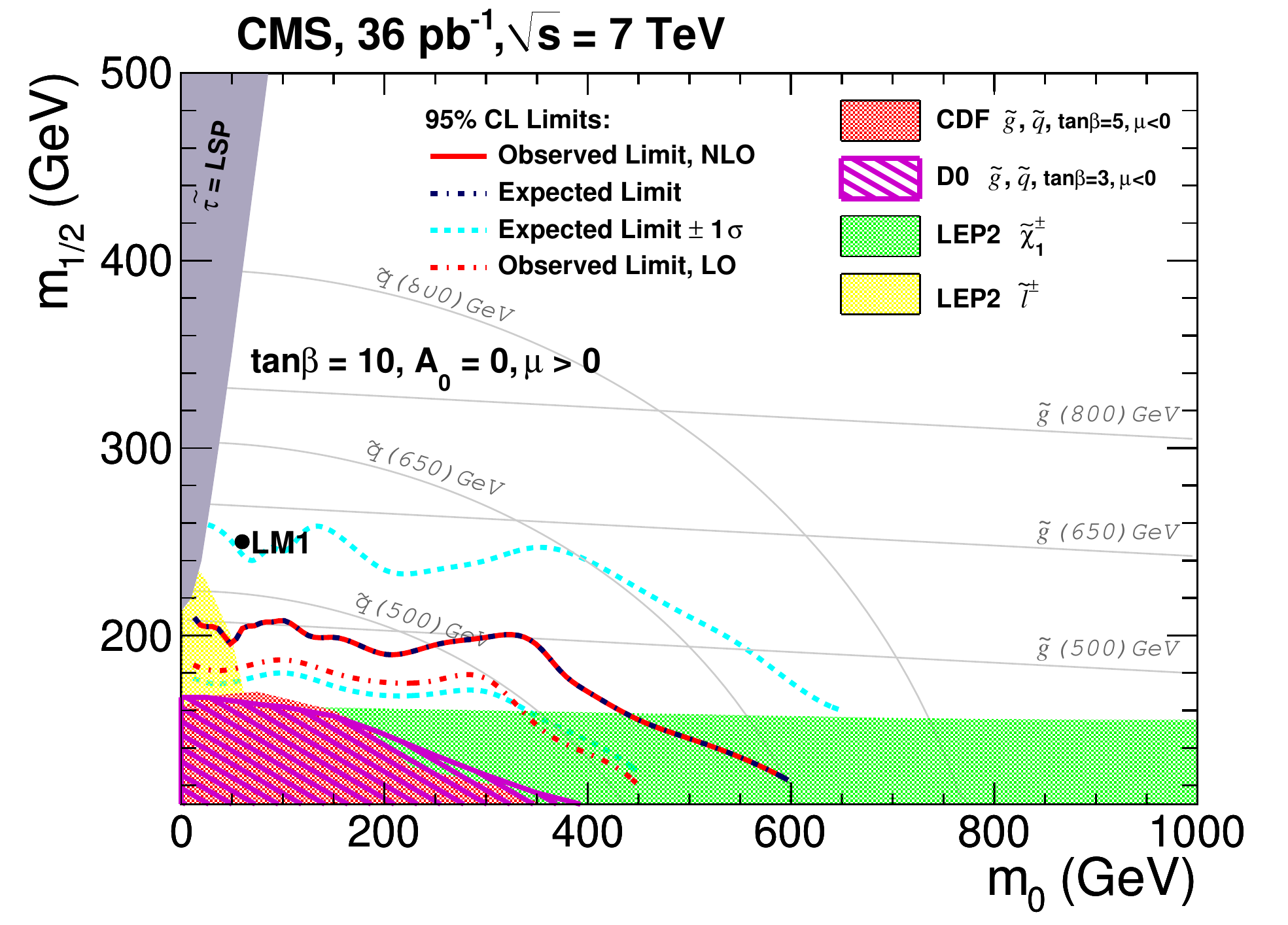} }
\end{center}
\caption{Exclusion region in the CMSSM $m_{1/2}$ vs.~$m_0$ plane for $\tan\beta=10$, based on the loose selection of the lepton-spectrum 
method, using the combined electron and muon samples. The observed limit is given for both LO and NLO assumed cross sections of the SUSY model points. In addition
to the observed limit, the expected limit under the assumption of no signal contribution and the
$\pm 1\sigma$ limits are shown.}\label{fig:ExclusionRegion_tanb10_loose}
\end{figure}
\begin{figure}[tbp!]
\begin{center}
\subfigure{\includegraphics[angle=0,height=0.55\textwidth]{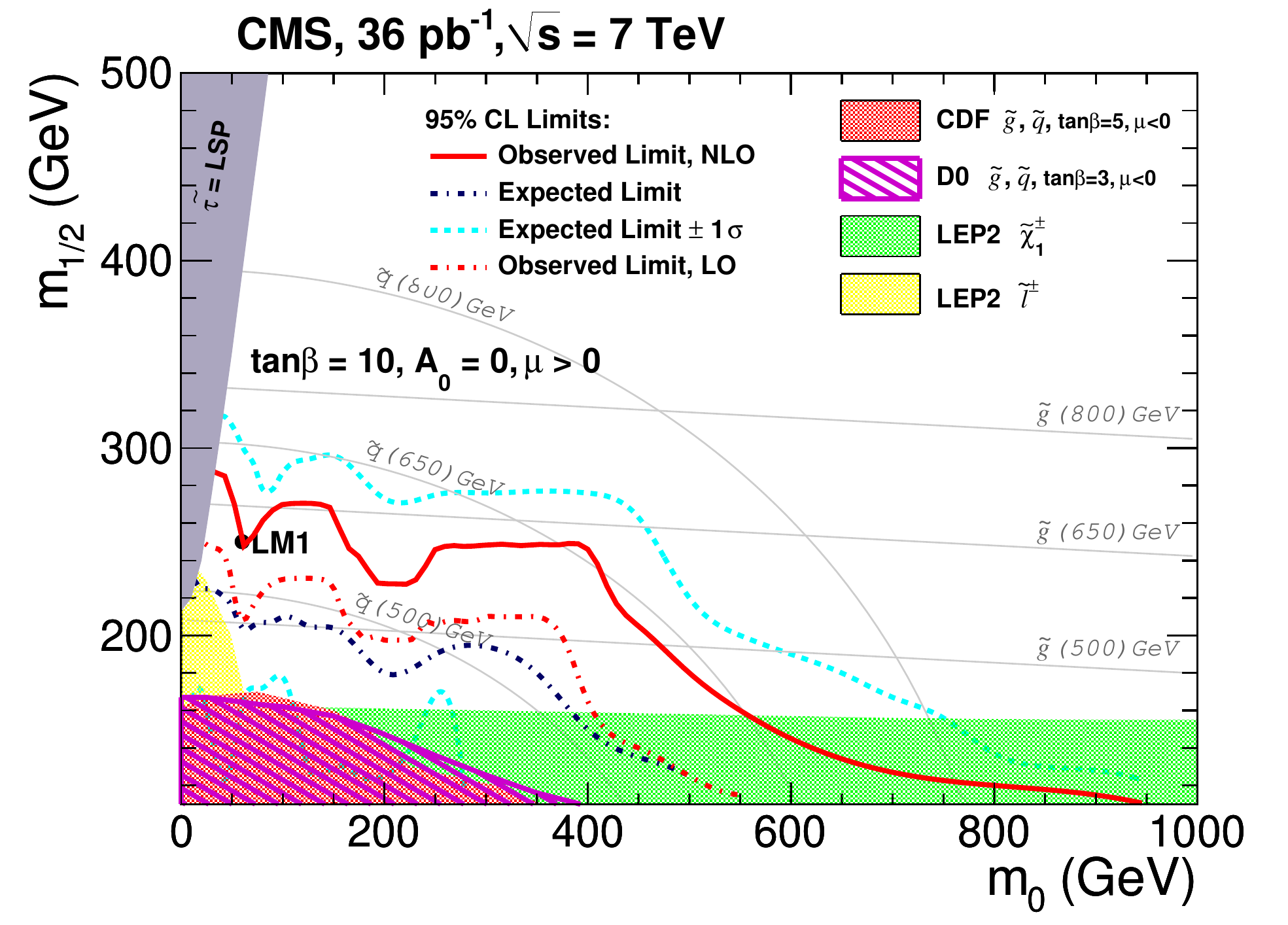} }
\end{center}
\caption{Exclusion region in the CMSSM $m_{1/2}$ vs.~$m_0$ plane for $\tan\beta=10$, based on the tight selection of the lepton-spectrum
method, using the combined electron and muon samples.}\label{fig:ExclusionRegion_tanb10_tight}
\end{figure}

To obtain a more comprehensive result, we perform scans of CMSSM models 
to determine whether a given set of parameters is excluded. 
The Monte Carlo samples are initially generated using leading-order 
cross sections; the predicted yields are corrected using 
process-dependent NLO cross sections evaluated with \PROSPINO~\cite{Beenakker:1996ch}. 
Figure~\ref{fig:ExclusionRegion_tanb10_loose} shows the 
limit curves resulting from the loose selection, 
evaluated in the $m_{1/2}$ vs.~$m_0$ plane, with
the values of the remaining CMSSM parameters fixed at 
$\tan\beta=10$, $A_0=0$, and $\mu>0$. The
corresponding curves for the tight selection, which exclude
a larger region,  are shown in 
Fig.~\ref{fig:ExclusionRegion_tanb10_tight}.
For reference, the plots include curves of constant gluino and 
squark masses. The lines of constant gluino mass
are approximately horizontal with $m(\PSg)\approx 2.5\, m_{1/2}$.
The lines of constant squark mass are
strongly curved in the $m_{1/2}$ vs.~$m_{0}$ plane. 
The total signal cross section decreases as a function of 
$m_{1/2}$ and $m_{0}$, roughly following the squark-mass contours.

The signal efficiency is defined for each model as the number of events passing the reconstructed-event selection, divided by the total number of SUSY events generated in the simulation,
summing over all decay chains. (This definition of efficiency therefore incorporates
the many different branching fractions leading to single-lepton final states, and
it also includes the loss in efficiency associated with the dilepton veto.)
The efficiency increases
with $m_{1/2}$ but is relatively uniform as a function of $m_{0}$.
In the tight selection, the efficiency in the combined \Pe\  and $\mu$ channels 
is roughly 2\% at $m_{0}$ = 250 GeV. For the benchmark model LM0 (LM1), the efficiency is 
3.2\% (3.6\%) in the loose selection and 0.6\% (1.7\%) in the 
tight selection. If one were simply to require a reconstructed \Pe\  or $\mu$ satisfying 
the acceptance requirements, the efficiency for LM1 would be 13\%. 

The exclusion plots show the observed limits, as well as the expected limits and the 
expected limits plus-or-minus one standard deviation ($\pm 1\sigma$). The loose selection
has a smaller $\pm 1\sigma$ band and in fact provides a stronger exclusion of the 
low-mass part of the range. The small dips in the 
exclusion limits in the $m_0$ range 50--200 GeV arise from 
corresponding dips in the efficiency curves; the falloff 
in the exclusion limits around $m_0$ = 350--400 GeV is due to
the decrease in the cross section. The tight selection excludes 
gluino masses below ${\approx} 600$ GeV for $m_0$ below 
${\approx}400$ GeV in the context of the CMSSM framework.

%% file: Conclusions_Final.tex
\section{Conclusions}
\label{sec:Conclusions}

Using a sample of proton-proton collisions at $\sqrt s=7$ TeV corresponding
to an integrated luminosity of 36 pb$^{-1}$, 
we have performed a search for new physics with the experimental signature
of at least four jets, an isolated, high-\pt lepton, and large missing transverse momentum. The
overall shapes of the kinematic distributions observed in data are consistent with expectations from SM simulated
event samples, indicating that the sample is dominated by \cPqt\cPaqt\ and \PW +jets events. 

To probe for new physics, control
samples in the data are used to predict the background contributions in the signal region.
The primary motivation for this approach is to avoid the direct use of 
simulated background event samples for predicting the 
extreme tails of kinematic distributions. 
The first background determination method focuses on the jets and $\met$. Using
the two-dimensional space of \HT and $\met/\sqrt{\HT}$, 
three control regions are defined in data, 
from which one predicts the SM background in the fourth region. This region has both
high \HT and high $\met/\sqrt{\HT}$ and hence is most sensitive to a signal
contribution. The observed event yield in the signal region is 
consistent with this prediction based on the control samples in the data.

The second method relies on the close relationship between two fundamental observables: the lepton \pt\
distribution and the $\met$ distribution, in the dominant SM background backgrounds
with a single isolated lepton. 
This connection arises from the fact
that the lepton and neutrino are produced together in the two-body decay of the \PW\  boson,
for both \cPqt\cPaqt\ and \PW +jets events. Smaller backgrounds from the feed-down of \cPqt\cPaqt\ 
dilepton events, from $\tau\to\ell$ decays in \cPqt\cPaqt\ or \PW +jets events, and from QCD multijet 
processes are also estimated from control samples in the data. In the muon channel, we
observe two events in the high-$\met$, high-\HT signal region (tight selection), 
as compared with $1.8\pm1.1\pm0.8$ 
SM events predicted; in the electron channel no events are observed, as compared with $1.4\pm0.9\pm0.5$
SM events predicted. The systematic uncertainties on the background predictions are correlated,
and the total background prediction is $3.2\pm 2.3$ events.
 
Finally, we interpret these results in the framework of the CMSSM, reporting exclusion
regions as a function of $m_{1/2}$ and $m_0$, for $\tan\beta=10$. The tight selection excludes
gluino masses below ${\approx} 600$ GeV for $m_0$ below ${\approx}400$ GeV 
in the context of the CMSSM framework.

%% file: Acknowledgments.tex
\section*{Acknowledgments}

We wish to congratulate our colleagues in the CERN accelerator departments for the excellent performance of the LHC machine. We thank the technical and administrative staff at CERN and other CMS institutes, and acknowledge support from: FMSR (Austria); FNRS and FWO (Belgium); CNPq, CAPES, FAPERJ, and FAPESP (Brazil); MES (Bulgaria); CERN; CAS, MoST, and NSFC (China); COLCIENCIAS (Colombia); MSES (Croatia); RPF (Cyprus); Academy of Sciences and NICPB (Estonia); Academy of Finland, MEC, and HIP (Finland); CEA and CNRS/IN2P3 (France); BMBF, DFG, and HGF (Germany); GSRT (Greece); OTKA and NKTH (Hungary); DAE and DST (India); IPM (Iran); SFI (Ireland); INFN (Italy); NRF and WCU (Korea); LAS (Lithuania); CINVESTAV, CONACYT, SEP, and UASLP-FAI (Mexico); MSI (New Zealand); PAEC (Pakistan); SCSR (Poland); FCT (Portugal); JINR (Armenia, Belarus, Georgia, Ukraine, Uzbekistan); MST, MAE and RFBR (Russia); MSTD (Serbia); MICINN and CPAN (Spain); Swiss Funding Agencies (Switzerland); NSC (Taipei); TUBITAK and TAEK (Turkey); STFC (United Kingdom); DOE and NSF (USA). Individuals have received support from the Marie-Curie programme and the European Research Council (European Union); the Leventis Foundation; the A. P. Sloan Foundation; the Alexander von Humboldt Foundation; the Associazione per lo Sviluppo Scientifico e Tecnologico del Piemonte (Italy); the Belgian Federal Science Policy Office; the Fonds pour la Formation \`a la Recherche dans l'Industrie et dans l'Agriculture (FRIA-Belgium); the Agentschap voor Innovatie door Wetenschap en Technologie (IWT-Belgium); and the Council of Science and Industrial Research, India. 

%% file: SUS-10-006-authorlist.tex
\textbf{Yerevan Physics Institute,  Yerevan,  Armenia}\\*[0pt]
S.~Chatrchyan, V.~Khachatryan, A.M.~Sirunyan, A.~Tumasyan
\vskip\cmsinstskip
\textbf{Institut f\"{u}r Hochenergiephysik der OeAW,  Wien,  Austria}\\*[0pt]
W.~Adam, T.~Bergauer, M.~Dragicevic, J.~Er\"{o}, C.~Fabjan, M.~Friedl, R.~Fr\"{u}hwirth, V.M.~Ghete, J.~Hammer\cmsAuthorMark{1}, S.~H\"{a}nsel, M.~Hoch, N.~H\"{o}rmann, J.~Hrubec, M.~Jeitler, W.~Kiesenhofer, M.~Krammer, D.~Liko, I.~Mikulec, M.~Pernicka, B.~Rahbaran, H.~Rohringer, R.~Sch\"{o}fbeck, J.~Strauss, A.~Taurok, F.~Teischinger, P.~Wagner, W.~Waltenberger, G.~Walzel, E.~Widl, C.-E.~Wulz
% ...
\vskip\cmsinstskip
% ...
1:~~Also at CERN, European Organization for Nuclear Research, Geneva, Switzerland\\
% ...
